\documentclass[twocolumn,twocolappendix]{aastex701}
\usepackage{natbib}
\begin{document}

\shorttitle{ALMA Review Process}
\shortauthors{Carpenter \& Corvill\'on}

\title{Distributed Peer Review at ALMA: An Empirical Comparison with Panel-Based Review}

\author[0000-0003-2251-0602]{John M. Carpenter}
\affiliation{Joint ALMA Observatory, Avenida Alonso de C\'ordova 3107, Vitacura, Santiago, Chile}
\email{john.carpenter@alma.cl}

\author{Andrea Corvill\'on}
\affiliation{Joint ALMA Observatory, Avenida Alonso de C\'ordova 3107, Vitacura, Santiago, Chile}
\email{andrea.corvillon@alma.cl}

\correspondingauthor{John M. Carpenter}
\email{john.carpenter@alma.cl}

\begin{abstract}

Large astronomical observatories are increasingly adopting distributed peer review (DPR) to manage growing proposal volumes, yet empirical comparisons with panel-based systems remain limited. Beginning in 2021 (Cycle~8), the Atacama Large Millimeter/submillimeter Array (ALMA) transitioned to DPR for the majority of proposals, with DPR applied to nearly all proposals from Cycle~9 onward. Under DPR, each Principal Investigator (PI) designates a reviewer to evaluate a set of proposals without collective discussion. This study analyzes over 20,000 proposals and 160,000 reviews spanning 13 cycles to assess the impact of this change. We find that DPR largely reproduces the systematic ranking trends observed in panel evaluations across PI demographics, technical characteristics, and scientific areas, consistent with panel outcomes both before and after discussion. Scientific diversity among the top-ranked proposals is similar between DPR and post-discussion panel rankings. Individual proposal ranks show substantial dispersion under both DPR and panel assessments prior to discussion, with discussion only partially reducing this variance. The observed dispersion therefore reflects intrinsic variation in reviewer judgments rather than a byproduct of the distributed process itself. In Cycle~12, reviewers rated the majority of DPR written comments as high or adequate quality, with no dependence on reviewer career stage. However, 10\% of reviews were rated as low quality, highlighting the challenge of maintaining quality standards across the $\sim$16,000 reviews produced each cycle. Overall, our results indicate that DPR reproduces the population-level ranking structure obtained in panel review, despite differences in review mechanics and the role of discussion.

\end{abstract}

\section{Introduction}
\label{sec:introduction}

Peer review faces well-documented challenges, including systematic biases \citep{Lee13, Wenneras97}, low inter-reviewer agreement \citep{Cole81, Pier18}, and growing scalability pressures as proposal volumes increase \citep{Merrifield09}. In astronomy, these challenges are amplified at major facilities experiencing sustained growth in demand. The Atacama Large Millimeter/submillimeter Array (ALMA) exemplifies this trend: since beginning operations in 2011, annual proposal submissions have increased from 919 in Cycle~0 to roughly 1700 in recent cycles.

During its first eight cycles (Cycles~0--7, covering the years 2011--2019), ALMA employed a traditional panel-based review system in which a fixed group of reviewers discussed a shared set of proposals to produce consensus rankings within a defined scientific scope. Each panelist evaluated approximately 100 proposals, with a substantial fraction subsequently discussed collectively to establish relative priorities. By Cycle~7, the system required 158 reviewers across 25 panels, a threefold increase relative to Cycle~0. This expansion increased the logistical complexity of organizing in-person meetings, imposed growing costs for international travel, and placed increasing demands on the community to provide reviewers willing to dedicate substantial time to the process.

To address these operational pressures, ALMA transitioned to a distributed peer review (DPR) process \citep{Merrifield09} beginning in Cycle~8 in 2021 (see \citealt{DonovanMeyer22}), following a small-scale pilot conducted in a supplemental call in Cycle~7. Under DPR, each Principal Investigator (PI) designates a member of the proposing team to serve as a reviewer. Each reviewer is assigned ten proposals to evaluate independently and remotely, without panel discussion or collective ranking. Participation is mandatory: failure to complete the assigned reviews results in automatic rejection of the reviewer's own proposal, regardless of scientific merit.

Several major observatories have adopted DPR-like approaches for limited proposal subsets. These include Gemini’s fast-turnaround proposals \citep{Mason14} and smaller programs at the European Southern Observatory \citep[ESO;][]{Patat19, Kerzendorf20, Jerabkova23, Jerabkova25}. Space-based observatories such as HST and JWST employ independent, non-discussion-based evaluations for some proposal categories, though reviewers are selected by the observatory rather than designated by proposing teams.

Following the introduction of DPR in Cycle~8, 
ALMA applied it to essentially its entire proposal portfolio beginning in Cycle~9, exempting only proposals designated as Large Programs, which account for $\sim$2\% of submitted proposals. This makes ALMA's transition a uniquely large-scale empirical test of distributed peer review within astronomy, both in terms of proposal volume ($\sim$1700 per cycle) and the fraction of the portfolio reviewed via DPR.

Traditional panel review and distributed peer review differ in how independent evaluations are combined. In panel review, reviewers first score proposals independently, after which a fixed group discusses a shared set of proposals and revises scores to produce consensus rankings. DPR retains independent evaluation but replaces structured panel discussion and rescoring with statistical aggregation of assessments from a much larger pool of reviewers, approximately 1000 reviewers per cycle compared to $\sim$150 under panels. In ALMA's implementation, DPR also permits participation by early-career members of proposing teams, including graduate students and postdoctoral researchers, whereas ALMA's panels were composed primarily of more senior researchers, thus expanding the range of career stages represented among reviewers. This shift raises questions about whether aggregation without discussion can achieve fairness and scientific breadth comparable to those of structured panel review, emphasizing broad community participation over consensus-building through deliberation.

While DPR-type processes have been widely adopted in fields such as computer science \citep{Shah22}, and applied in limited contexts within grant funding \citep{RoRI25} and astronomy \citep[e.g.,][]{Patat19, Jerabkova23, Jerabkova25}, comprehensive empirical comparisons of DPR outcomes against traditional panel review remain scarce. ALMA is uniquely positioned to address this gap. Its 13-cycle dataset, consisting of eight cycles of panel review and five of DPR (Cycles~8--12), comprises over 20,000 proposals, allowing for a systematic comparison of outcomes under both evaluation structures. The transition to DPR coincided with the introduction of dual-anonymous review (where both proposers’ and reviewers’ identities are concealed), making it impossible to isolate the independent effect of each change. Nevertheless, the dataset’s size and temporal span allow for the identification of broad patterns in ranking behavior. With oversubscription rates exceeding 7:1, even modest shifts in ranking can substantially influence which science is ultimately pursued, underscoring the practical significance of detecting systematic effects.

Previous studies have analyzed ALMA's panel-based system for Cycles~0--6 \citep[see also \citealt{Lonsdale16}]{Carpenter20}, while \citet{DonovanMeyer22} and \citet{Carpenter22} provided initial evaluations of the Cycle~8 DPR implementation. These early DPR analyses were necessarily limited by the availability of only a single DPR cycle. This paper builds upon that foundation using a dataset spanning all 13 ALMA cycles and more than 160,000 individual reviews, representing one of the largest empirical comparisons of distributed peer review and panel-based review to date. By examining five completed DPR cycles alongside eight panel cycles, we assess not only ranking outcomes but also systematic trends, scientific diversity, reviewer disagreement, and review quality. Specifically, this paper addresses four core questions:

\begin{enumerate}
\item Does DPR reproduce or alter systematic demographic and technical trends in proposal rankings?
\item Does independent evaluation affect the scientific diversity of top-ranked proposals?
\item Does reviewer disagreement under DPR differ from the variance inherent in panel evaluations?
\item Does review quality under DPR meet expected standards?
\end{enumerate}

The paper is organized as follows. Section~\ref{sec:data} describes the proposal and review data used in this analysis. Section~\ref{sec:systematics} compares systematic trends in proposal rankings between panel and DPR cycles across multiple dimensions. Section~\ref{sec:diversity} examines the diversity of scientific topics among top-ranked proposals. Section~\ref{sec:ranks} analyzes the dispersion, distribution, and pairwise agreement of individual proposal rankings. Section~\ref{sec:quality} evaluates the quality of the reviews. We discuss the implications of our findings in Section~\ref{sec:discussion} and summarize our main conclusions in Section~\ref{sec:conclusions}.

\section{Data}
\label{sec:data}

This analysis uses proposal and review data from ALMA Cycles~0--12, spanning 2011--2025 and encompassing over 20,000 proposals and more than 160,000 individual reviews. We focus on proposals submitted to the main call, excluding Large Programs (which follow a separate review process), Director’s Discretionary Time proposals, and supplemental calls.

ALMA employed panel-based review for Cycles~0--7 (2011--2019). Beginning in Cycle~8 (2021), DPR was introduced for most proposals, while medium proposals and Large Programs continued to be reviewed by panels in that cycle. From Cycle~9 onward, DPR has been used for all main-call proposals except Large Programs. Table~\ref{table:data} summarizes the number of proposals, reviewers, and individual reviews in each cycle under the two review systems. Cycle~8 represents a transitional year in which panel review and DPR were both used for different subsets of proposals; because these subsets differ systematically in size and scope, the two processes are listed separately in the table.

\subsection{Panel Review Process}
\label{sec:data:panels}

The ALMA panel review process has been described in detail by \citet{Carpenter20}; here we summarize the aspects relevant to this analysis. Under the panel-based system, ALMA organized proposals into topical panels of 6--8 reviewers, with each panel evaluating between 61 and 163 proposals per cycle (mean 97, median 94) within a broadly defined scientific area. Proposals were assigned based on the scientific category selected by the PI at submission: 1) Cosmology and the high-redshift universe, 2) Galaxies and galactic nuclei, 3) Interstellar medium, star formation, and astrochemistry, 4) Circumstellar disks, exoplanets, and the Solar System, and 5) Stellar evolution and the Sun.

The panel review proceeded in two stages. In Stage~1, each reviewer independently assessed all proposals for which they had no conflict of interest, providing a numerical score\footnote{In the actual review process conducted by the JAO, individual reviewer Stage 1 scores were normalized to have the same mean and standard deviation before averaging. This normalization is not applied in our analysis to maintain consistency between Stage 1 and Stage 2 treatment; see \citet{Carpenter20}.} and written comments. Following Stage~1, the lowest-ranked proposals were triaged and removed from further consideration on a region-dependent basis, with triage levels adjusted to maintain at least a factor of two oversubscription in requested observing time relative to available time for each region; this typically resulted in approximately 30\% of proposals being triaged overall. The remaining proposals then proceeded to Stage 2 in-person meetings, during which proposals were discussed and rescored.

For panel-reviewed cycles (Cycles~0--7), proposals within a panel were ranked 1 (best) to N (worst) based on average reviewer scores at each stage; these panel-level rankings were then normalized by dividing by the number of proposals in each panel to yield normalized rankings between 0 (best) and 1 (worst). Because panels were organized by scientific category, this normalization effectively accounts for differences in both panel size and scoring standards across scientific areas. Proposals from all panels were merged into a single cycle-wide ranked list by sorting these normalized rankings, with ties broken randomly \citep{Carpenter20}.

In Cycles~0--6, the review process was single-anonymous: the PI and co-investigators were listed on the proposal cover sheet, while the identity of the reviewers was not revealed. In Cycle~7, ALMA introduced partial anonymization by randomizing the order of the proposal team list without identifying the PI. During the subsequent transition in Cycle~8, a limited panel-based review was retained for a small subset of proposals. Because this subset is not representative of the full proposal population, panel results from Cycle~8 are excluded from the panel comparison analysis.

Panel reviewer composition was explicitly balanced to reflect the fraction of ALMA observing time assigned to each region, with nominal targets of 33.75\% for Europe, 33.75\% for North America, 22.5\% for East Asia, and 10\% for Chile. Reviewers from all remaining regions (1--5\%) are grouped as ``Other".\footnote{In this paper, Europe, North America, and East Asia refer to the ALMA executive partners. ``Other" includes reviewers affiliated with countries outside these partner regions and Chile.} This approach ensures that the distribution of reviewer expertise broadly mirrors the regional distribution of ALMA observing time.

\subsection{Distributed Peer Review Process}

The ALMA DPR process is described in detail in \citet{DonovanMeyer22}; here we summarize the key elements relevant to this study. A small-scale pilot of DPR was conducted in a supplemental call in Cycle~7 \citep{Carpenter20b}, allowing the community to become familiar with the process before its broader implementation in Cycle~8. Under DPR, each PI designates a member of the proposing team to serve as a reviewer. Reviewers are assigned 10 proposals using an algorithm designed to match proposals to reviewer expertise while ensuring that each proposal receives exactly 10 reviews. Individual reviewers may receive multiple proposal sets, so the number of distinct reviewers is typically smaller than the number of proposals. In Cycles~8 and 9, assignments were based on keyword matching between submitted proposals and self-reported reviewer expertise \citep{DonovanMeyer22}. Beginning in Cycle~10, ALMA adopted a machine-learning approach that infers reviewer expertise from current and prior proposal submissions and uses optimization algorithms to improve assignment accuracy \citep{Carpenter25}. DPR has employed dual-anonymous review since Cycle~8, with reviewers not provided the identities of proposal team members.

The review proceeds in two stages. In Stage~1, reviewers independently rank their 10 assigned proposals from 1 (best) to 10 (worst) and provide written comments; completion of Stage~1 is mandatory, and failure to submit reviews results in automatic rejection of the reviewer’s proposal. In Stage~2, reviewers may view anonymized comments from other reviewers for the same proposals and revise their ranks and comments. Only 6.5\% of individual rankings were modified during Stage~2, with modest impact on the final aggregated rankings (see Appendix~\ref{app:stage12_comparison_dpr}). For all analyses in this paper, we use the final Stage~2 ranks.

Final proposal rankings are obtained by aggregating individual Stage~2 ranks. In Cycles~8--9, rankings were combined without explicit normalization by science category. Beginning in Cycle~10, rankings were normalized by science category to be consistent with the treatment of panel-based rankings. No systematic category-level biases were observed in Cycles~8--9, but normalization has been applied in all subsequent cycles to ensure fairness across science categories.

Unlike panel review, DPR draws from a global pool of expert reviewers without explicit regional balancing. Across DPR cycles (Cycles~8--12), the reviewer composition has differed from panel-era targets, with European reviewers comprising 37--41\% of the pool, North American reviewers 27--29\%, East Asian reviewers 20--22\%, Chilean reviewers 3--6\%, and Other 6--8\%. Compared to panel-era composition (see Section~\ref{sec:data:panels}), European reviewers tend to be over-represented in DPR while Chilean and North American reviewers tend to be under-represented.

\begin{deluxetable}{cccrrr}
\tabletypesize{\footnotesize}
\tablewidth{0pt}
\tablecaption{Summary of Reviews per Cycle \label{table:data}}

\tablehead{
\colhead{Cycle} &
\colhead{Year} &
\colhead{Process} &
\colhead{Proposals} &
\colhead{Reviewers} &
\colhead{Reviews}
}
\startdata
0 & 2011 & Panels & 919 & 49 & 3,667 \\
1 & 2012 & Panels & 1,131 & 77 & 6,588 \\
2 & 2013 & Panels & 1,381 & 77 & 8,135 \\
3 & 2015 & Panels & 1,578 & 96 & 10,470 \\
4 & 2016 & Panels & 1,544 & 144 & 11,909 \\
5 & 2017 & Panels & 1,639 & 146 & 12,632 \\
6 & 2018 & Panels & 1,818 & 146 & 14,107 \\
7 & 2019 & Panels & 1,759 & 158 & 10,702 \\
8 & 2021 & Panels & 198 & 39 & 1,266 \\
8 & 2021 & DPR & 1,496 & 1,015 & 14,959 \\
9 & 2022 & DPR & 1,707 & 1,085 & 17,070 \\
10 & 2023 & DPR & 1,634 & 1,097 & 16,328 \\
11 & 2024 & DPR & 1,654 & 1,089 & 16,526 \\
12 & 2025 & DPR & 1,595 & 1,106 & 15,931 \\
Total &  & Panels & 11,967 & 932 & 79,476 \\
Total &  & DPR & 8,086 & 5,392 & 80,814 \\
\enddata
\tablecomments{
The data exclude Large Programs, Director’s Discretionary Time proposals, and supplemental calls. 
Cycle~8 included both panel-reviewed and DPR-reviewed proposals, which are listed separately.
Reviewer counts represent unique reviewers within each cycle; the same individual may appear in multiple cycles.
}
\end{deluxetable}

\section{Systematic Trends in Proposal Rankings}
\label{sec:systematics}

A central question in evaluating the transition to DPR is whether proposal ranking outcomes differ systematically from those produced by the traditional panel-based process. Here we compare the distribution of proposal rankings under panel reviews (Cycles~0--7) and DPR (Cycles~8--12) across multiple dimensions, including PI demographics (regional affiliation, gender, and experience), technical characteristics (observing modes, receiver bands, and requested observing time), and special proposal categories. Our goal is to assess whether DPR reproduces the broad, community-wide evaluation patterns observed under panel reviews, or whether the change in review process has systematically advantaged or disadvantaged particular classes of proposals or PIs.

To enable these comparisons, proposal rankings are converted to cumulative distribution functions (CDFs) and normalized within each cycle such that a value of 0 corresponds to the best-ranked proposal and 1 to the worst-ranked proposal in that cycle. Shaded bands around each CDF indicate the $\pm 1\sigma$ (68\%) confidence intervals for the empirical distribution, computed from exact beta-distribution confidence bounds for sample quantiles. Differences between ranking distributions are assessed using the Anderson--Darling
$k$-sample test \citep{Scholz19}, which provides greater sensitivity to distributional differences across all ranks than the Kolmogorov-Smirnov test. Throughout this section, we report $p$-values from this test, with smaller values indicating stronger evidence that two samples are drawn from different parent distributions. We adopt a significance threshold of $\alpha=0.01$. Because multiple comparisons are performed within each subsection, we apply the Benjamini--Hochberg procedure to control the false discovery rate at $\alpha=0.01$. The correction is applied separately within each subsection, as each addresses a distinct set of related hypotheses. All $p$-values are reported so that readers can assess the strength of evidence for individual comparisons.

Ideally, panel review would be compared to DPR using Stage~2 panel rankings, since discussion reconciles individual assessments into a consensus. However, following Stage~1, the lowest-ranked proposals are removed through region-dependent triage to focus deliberation on the remaining submissions (see Section~\ref{sec:data:panels}). As a result, Stage~2 rankings apply only to a selected subset, and this truncation can distort subgroup statistics (e.g., by region, gender, or career stage) when these characteristics vary across regions. To enable a fair comparison with DPR across the full proposal population, we therefore use Stage~1 panel rankings in this section, reflecting the initial assessments of all submissions. Appendix~\ref{app:stage12_comparison_panels} confirms that macro-level systematic patterns are largely preserved between Stage~1 and Stage~2, indicating that Stage~1 rankings faithfully reflect the broad outcomes of the panel process.

Data are aggregated across all cycles within each review system to emphasize persistent systematic patterns rather than cycle-to-cycle fluctuations. In the following subsections, we examine each dimension in turn, beginning with PI demographics (Sections~\ref{subsec:region}--\ref{subsec:experience}), followed by technical and observational characteristics (Sections~\ref{subsec:modes}--\ref{subsec:time12m}).

Because the transition to DPR was not a controlled experiment, observed differences between panel and DPR outcomes cannot be uniquely attributed to the change in review process. Other contemporaneous factors, such as evolving scientific priorities, the introduction of dual-anonymous review, and broader changes in the research landscape, may also contribute to differences observed across cycles. This section highlights similarities and differences between panel and DPR rankings, while the interpretation of any differences is discussed in Section~\ref{sec:discussion}.

\subsection{PI Regional Affiliation}
\label{subsec:region}

\begin{figure}
\centering
\includegraphics[width=\columnwidth]{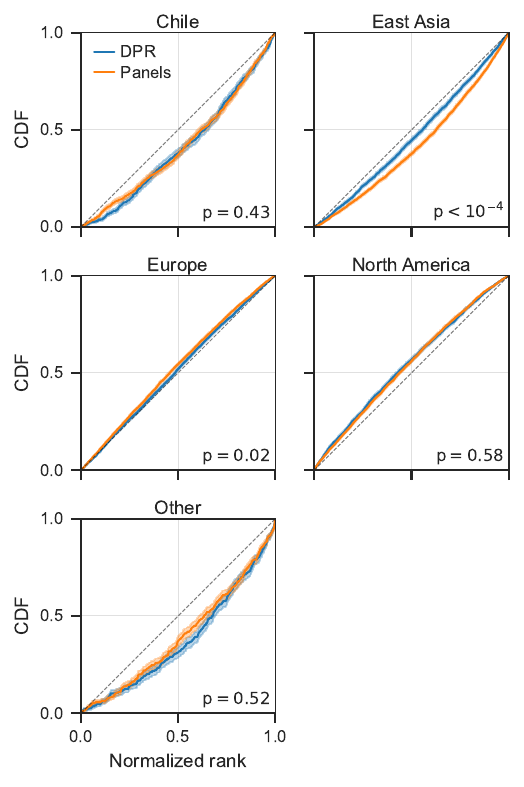}
\caption{Cumulative distribution functions of normalized proposal ranks in Stage~1 panel reviews in Cycles~0--7 and the final DPR ranks in Cycles~8--12, grouped by PI regional affiliation. The $p$-value in each panel indicates the result of a two-sample Anderson--Darling test comparing the panel and DPR rank distributions. The dashed line represents the cumulative distribution expected for uniformly distributed ranks.}
\label{fig:panels_dpr_region}
\end{figure}

Figure~\ref{fig:panels_dpr_region} compares the cumulative distributions of Stage~1 panel ranks from Cycles~0--7 and the final DPR ranks in Cycles~8--12, grouped by the regional affiliation of the proposal's PI. In these figures, the dashed line shows the cumulative distribution expected for uniformly distributed ranks; curves shifted above this line indicate systematically better-than-average rankings, while curves shifted below indicate worse-than-average rankings. Proposals from North American and European PIs consistently receive above-average rankings in both panel reviews and DPR, with their cumulative distributions shifted toward better (lower numerical) ranks relative to the uniform distribution. In contrast, proposals from Chilean, East Asian, and Other regional PIs tend to receive below-average rankings in both review systems.

While the qualitative regional trends are similar for panels and DPR, proposals from East Asian PIs show a statistically significant shift toward better ranks under DPR relative to panels ($p<10^{-4}$). Examination by individual cycle shows this shift is absent in Cycle~8, first appears in Cycle~9, and strengthens through Cycle~12. Conversely, proposals from European PIs exhibit slightly poorer rankings under DPR than under panels ($p=0.02$); however, this difference is not significant after controlling the false discovery rate within the regional comparison set.

\subsection{PI Gender}
\label{subsec:gender}

Following the methodology of previous studies \citep{Lonsdale16, Carpenter20, Carpenter22}, the gender of each PI was inferred using publicly available information (e.g., institutional and personal webpages). In a small number of cases, gender information was obtained from ALMA user profiles, where PIs may voluntarily provide this information. Therefore, this classification is based primarily on how PIs present publicly and may not reflect self-identified gender in all cases. This analysis is limited to binary gender classifications (male and female), as these were the only categories that could be consistently inferred from available data. PIs for whom gender could not be clearly established, as well as those who self-identified as nonbinary or other in their ALMA user profiles, were excluded. 

Under dual-anonymous review, reviewers do not have access to PI identities, so gender cannot directly influence reviewer behavior except in cases where a reviewer believes they can infer the PI. However, indirect signals correlated with gender, such as writing style or choice of scientific topic, could potentially influence evaluations even in the absence of explicit identification.

\begin{figure}
\centering
\includegraphics[width=\columnwidth]{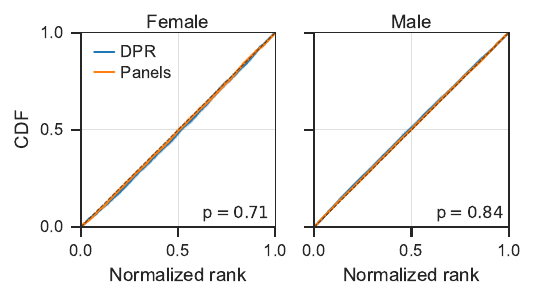}
\caption{Same as Figure~\ref{fig:panels_dpr_region}, but grouped by PI gender.}
\label{fig:panels_dpr_gender}
\end{figure}

Figure~\ref{fig:panels_dpr_gender} compares the cumulative distributions of Stage~1 panel ranks from Cycles~0--7 and the final DPR ranks in Cycles~8--12, grouped by the gender of the proposal PI. In both review systems, the rank distributions for proposals led by female and male PIs are similar and lie close to the uniform distribution. We find no statistically significant gender-dependent differences in proposal ranking outcomes between the two review systems, suggesting that any gender-dependent changes introduced by the transition are below the sensitivity of the present dataset.

\subsection{PI Experience}
\label{subsec:experience}

We investigate how proposal rankings vary with PI experience, defined as the number of ALMA cycles in which an individual has submitted at least one proposal as PI (not as co-investigator) up to and including a given cycle. Experience categories are evaluated independently in each cycle, so a PI may belong to different experience bins in different cycles. To highlight the strongest systematic effects identified in previous analysis \citep{Carpenter20}, we focus on four mutually exclusive categories: first-time PIs (submitting in only one cycle), second-time PIs (submitting in two cycles), third-time PIs (submitting in three cycles), and most experienced PIs (submitting proposals in every cycle from Cycle~0 up to and including a given cycle). PIs with intermediate levels of experience (i.e., those who have submitted in more than three cycles but not in every cycle) are excluded from this analysis. 

For panel reviews, we have excluded Cycle~7 data where the proposal team list was randomized, which was found to affect systematics related to PI experience \citep{Carpenter22}. This ensures a clean comparison between panel review under single-anonymous conditions and DPR under dual-anonymous review.

Because the most experienced category is defined dynamically on a per-cycle basis, lower-experience bins are introduced only once they can be defined without overlap. First-time PIs are therefore considered starting in Cycle~1, since all PIs in Cycle~0 are by definition first-time. Second-time PIs are defined beginning in Cycle~2, and third-time PIs in Cycle~3. As a result, the cycles contributing to each experience bin differ; for example, first-time PI data are drawn from Cycles~1--6, while third-time PI data are drawn from Cycles~3--6. This dynamic binning ensures consistent and non-overlapping definitions of PI experience across the transition from single-anonymous panel review to dual-anonymous DPR.

\begin{figure}
\centering
\includegraphics[width=\columnwidth]{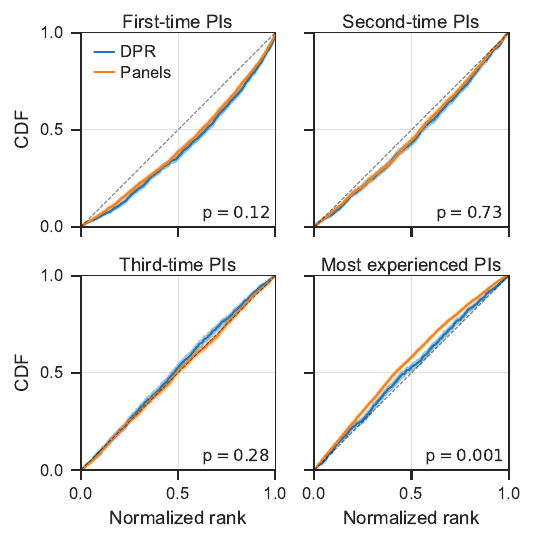}
\caption{Same as Figure~\ref{fig:panels_dpr_region}, but grouped by PI experience. Cycle~7 panel data were excluded (see text).}
\label{fig:panels_dpr_experience}
\end{figure}

The results, shown in Figure~\ref{fig:panels_dpr_experience}, indicate that ranking outcomes vary with PI experience in both review systems. First- and second-time PIs tend to receive lower-than-average rankings, while third-time PIs have rankings consistent with a uniform distribution. The most pronounced difference between the systems occurs among the most experienced PIs: this group ranked above average under panel review, but their advantage is significantly reduced under DPR ($p=0.001$). Notably, the improvement for second-time PIs reported by \citet{Carpenter22} following the introduction of randomization of investigator names in Cycle~7 does not persist when considering subsequent cycles ($p=0.73$).

\subsection{Observing Modes and Special Categories}
\label{subsec:modes}

\begin{figure}
\centering
\includegraphics[width=\columnwidth]{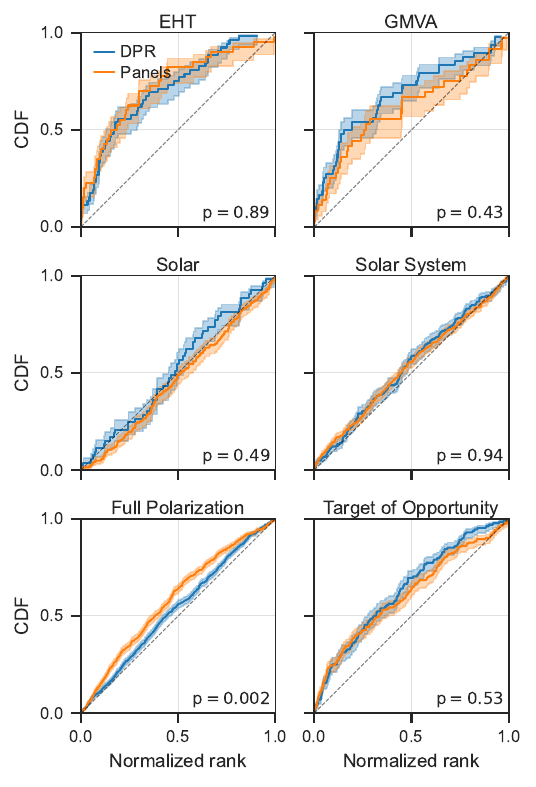}
\caption{Same as Figure~\ref{fig:panels_dpr_region}, but grouped by observing modes, science areas, and proposal type.}
\label{fig:panels_dpr_modes}
\end{figure}

Figure~\ref{fig:panels_dpr_modes} compares the cumulative distributions of proposal rankings for various observing modes, science areas, and special proposal types under panel reviews and DPR. For most categories, ranking behavior is qualitatively similar across the two review systems. Proposals requesting very long baseline interferometry (VLBI), including Event Horizon Telescope (EHT) and Global Millimeter VLBI Array (GMVA) observations, and Target of Opportunity observations receive above-average rankings relative to a uniform distribution. Proposals targeting Solar System objects (planets, moons, comets, and asteroids, excluding the Sun) also show mildly above-average rankings, while Solar proposals rank near the average.

The only statistically significant difference between panel and DPR outcomes is for full-polarization proposals, which receive higher rankings during panel reviews than during DPR ($p=0.0016$). 
Examining cycle-by-cycle ranking distributions shows that the aggregate panel-era advantage for full-polarization proposals is driven primarily by early cycles (Cycles~2--6), where the effect is strongest ($p = 2\times10^{-4}$ to $0.03$). The difference weakens in Cycle~7 ($p = 0.34$) and is not consistently observed across the DPR cycles (Cycles~8--12, $p = 0.04$–$0.49$). Despite this temporal variation, polarization proposals remain above average in both review systems.

\subsection{Receiver Band}
\label{subsec:band}

\begin{figure}
\centering
\includegraphics[width=\columnwidth]{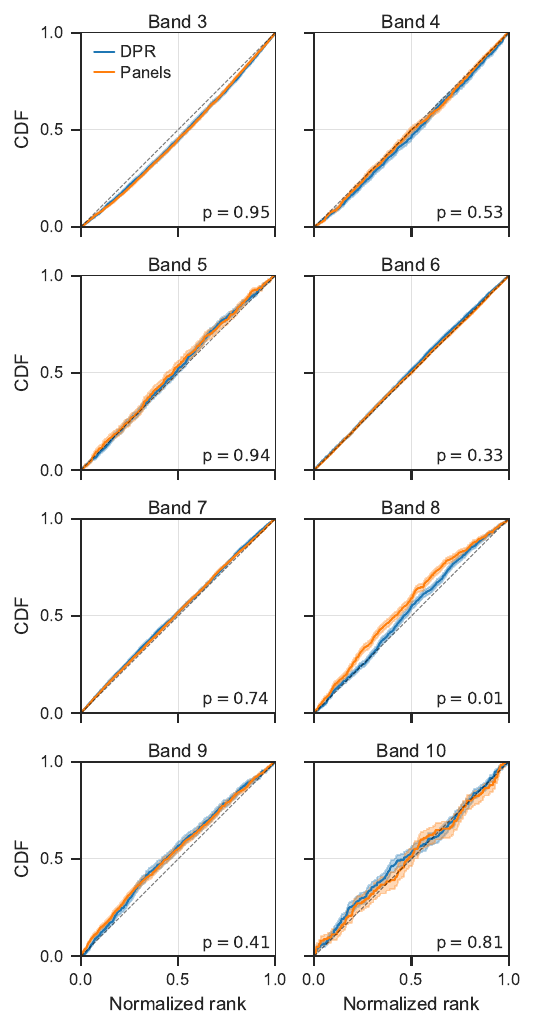}
\caption{Same as Figure~\ref{fig:panels_dpr_region}, but grouped by receiver band.}
\label{fig:panels_dpr_bands}
\end{figure}

Figure~\ref{fig:panels_dpr_bands} compares the cumulative distributions of proposal rankings by receiver band for panel reviews and DPR. Proposals requesting more than one receiver band appear in the plot for each requested band. Overall, the ranking distributions by band are similar across both review systems. Proposals requesting Band~3 tend to receive below-average rankings relative to a uniform distribution, whereas proposals requesting Bands~8 and 9 rank above average in both the panel and DPR processes.

The largest difference between the two systems is observed for Band~8, where proposals received higher rankings under panel review than under DPR ($p=0.01$); however, this difference is not significant after controlling the false discovery rate within the receiver band comparison set. Examining individual cycles, Band~8 proposals received above-average rankings in Cycles~2 and 3 ($p \simeq 0.01$--$0.03$), the first two cycles in which the capability was offered. From Cycle~4 onward, we do not detect statistically significant differences between the ranking distributions ($p=0.31$).

\subsection{Requested 12-m Array Time}
\label{subsec:time12m}

\begin{figure}
\centering
\includegraphics[width=\columnwidth]{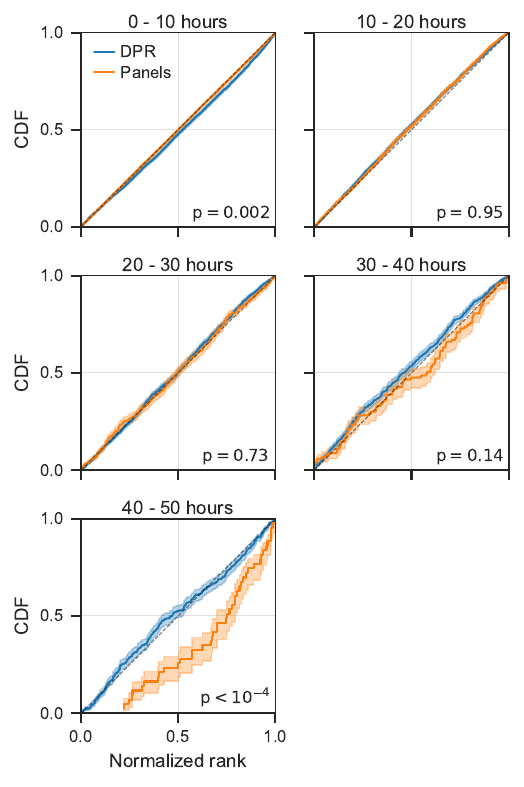}
\caption{Same as Figure~\ref{fig:panels_dpr_region}, but grouped by requested time on the 12-m Array.}
\label{fig:panels_dpr_time12m}
\end{figure}

Figure~\ref{fig:panels_dpr_time12m} compares the cumulative distributions of proposal rankings grouped by the requested observing time on the 12-m Array. Distinct differences between panel and DPR outcomes emerge at the shortest and longest requested time scales, while intermediate time requests show no statistically significant differences.

For proposals requesting less than 10~h, DPR rankings are shifted toward poorer outcomes relative to panels ($p=0.002$). In contrast, proposals requesting 10--20~h and 20--30~h show no significant differences between panels and DPR; in both review systems, the rank distributions for these categories are consistent with a uniform distribution.

For proposals requesting 30--40~h, panel-reviewed proposals tend to receive lower rankings than those reviewed under DPR, although this difference is not statistically significant ($p=0.14$). For the largest proposals (40--50~h), the ranking distributions differ significantly, with proposals reviewed under DPR receiving more favorable rankings than under panels ($p < 10^{-4}$).

Overall, these results indicate that differences between panel and DPR outcomes are most pronounced at the extremes of the requested time distribution, while proposals requesting intermediate amounts of observing time exhibit similar ranking behavior across the two review systems. We note that the trends observed at the largest requested times coincide with a broader shift in observatory guidance toward encouraging fewer, more ambitious programs in later cycles; this context is discussed in Section~\ref{subsec:discussion:obstime}.

\section{Scientific Diversity of Top-Ranked Proposals}
\label{sec:diversity}

In DPR, each reviewer evaluates a relatively small number of proposals independently, whereas panel reviews involve a larger shared proposal set and a collective discussion phase. This structural difference raises the question of whether independent evaluation (DPR) versus collective deliberation (panels) yields systematically different scientific breadth among highly ranked proposals. In this section, we compare the scientific diversity of top-ranked proposals under panel review and DPR to assess the impact of this shift. Throughout this section, ``scientific diversity" is defined operationally as the breadth of scientific areas represented among highly ranked proposals. This analysis focuses on categorical representation as a proxy for scientific breadth and does not address other dimensions of diversity.

To isolate the outcome of the peer review process itself, we focus on the top-ranked proposals produced by each review process rather than on the final accepted programs. Acceptance decisions are influenced by factors external to the peer review ranking, such as array configuration availability, regional balance, and weather conditions. Importantly, the typical requested observing time per proposal has grown substantially over the history of the observatory. Focusing on 12-m array requests excluding Large Programs (introduced in Cycle~4), the median requested time nearly tripled from Cycle~4 to Cycle~12. This increase has significantly outpaced the growth in available observing time, which rose more modestly from approximately 3000 to 4300 hours over a similar period.

As a consequence, the number of programs that can be scheduled has declined markedly: an average of $\simeq 420$ proposals were accepted in Cycles~4--7, compared to $\simeq 250$ in Cycles~8--12. Although this decline coincides temporally with the transition to DPR, its timing is consistent with the cumulative effects of increasing proposal size rather than a change in review mode. Because diversity metrics based on categorical representation are inherently sensitive to sample size, comparisons based on accepted proposals would confound reviewer behavior with evolving scheduling constraints unrelated to scientific evaluation. By instead selecting a fixed percentile of the ranked list, which yields a consistent sample size given that total submissions have remained stable within $\pm 8\%$ since Cycle~3, we enable a controlled comparison of review outcomes that isolates reviewer behavior from these evolving observatory constraints.

With this ranking-based selection framework in place, we assess scientific diversity using two complementary approaches: (1) an analysis of the predefined ALMA science keywords assigned by proposers, and (2) unsupervised topic modeling of proposal abstracts using Latent Dirichlet Allocation (LDA), the same methodology employed in the ALMA proposal assignment system starting in Cycle~10 \citep{Carpenter25}. For panel-reviewed cycles, we use the final post–Stage~2 rankings, as this section concerns the composition of the finalized top-ranked proposal set. Cycle~0 is excluded, as science keywords were not introduced until Cycle~1. We first examine diversity as defined by proposer-selected science keywords, followed by an analysis based on topic modeling of proposal abstracts.

Under the panel-based review system, rankings were normalized across panels to maintain proportional representation of the five broad science categories. In contrast, explicit category-level normalization was not applied during the initial DPR implementation (Cycles~8--9), and was introduced only in Cycles~10--12. The implications of this procedural difference for the results presented here are discussed further in Section~\ref{subsec:discussion:diversity}.

\begin{figure*}
\centering
\includegraphics[width=\textwidth]{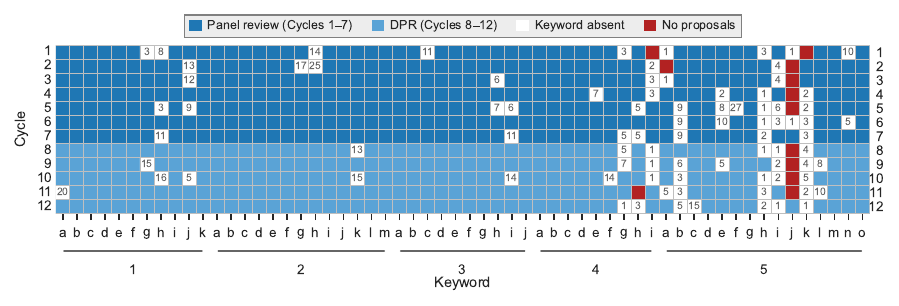}
\caption{
Representation of the 58 ALMA science keywords (defined in \citealt{Privon25}) among the top 15\% of proposals across Cycles~1--12. Each cell shows whether a given keyword appears among the top-ranked proposals in a given cycle. Blue squares indicate keywords represented by at least one proposal in the top 15\%, with darker and lighter blue corresponding to panel-based review (Cycles~0–7) and distributed peer review (Cycles~8–12), respectively. White squares indicate that proposals with that keyword were submitted but none ranked in the top 15\%; the number shown gives the total submissions for that keyword in that cycle. Red squares indicate that no proposals were submitted. Keywords are ordered by major science category (1–5) along the horizontal axis.
}
\label{fig:diversity_matrix}
\end{figure*}

\begin{figure}
\centering
\includegraphics[width=\columnwidth]{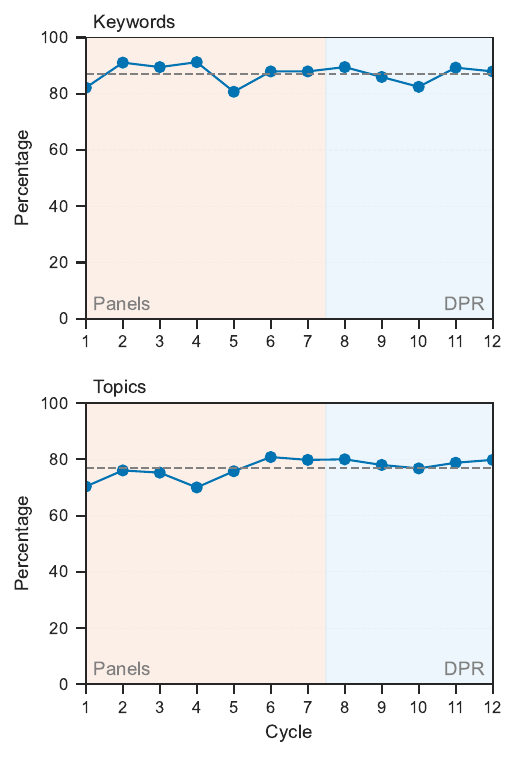}
\caption{
Scientific diversity of top-ranked proposals as a function of cycle, quantified by the fraction of ALMA science keywords represented (top) and the fraction of latent topics inferred from LDA topic modeling (bottom). In both panels, the top 15\% of ranked proposals in each cycle are analyzed. Shaded regions distinguish panel-review cycles (beige) from DPR cycles (blue). Horizontal dashed lines indicate the mean fraction represented across all cycles.
}
\label{fig:diversity}
\end{figure}

\subsection{Analysis Using ALMA Keywords}

We quantify scientific diversity by examining which ALMA science keywords are represented among the top-ranked proposals. Specifically, we consider the top 15\% of proposals, a threshold chosen to approximate the typical acceptance rate given recent oversubscription levels of $\sim$7:1. ALMA defines 58 science keywords spanning five broad science categories, and each proposal is assigned one or two keywords by the PI. Keywords are labeled by a numerical category (1--5) corresponding to major science areas, with letters indicating specific subtopics; see \citet{Privon25} for the full list of ALMA keywords.

Figure~\ref{fig:diversity_matrix} shows the representation of keywords among top-ranked proposals for each cycle. Blue squares indicate that at least one proposal with the given keyword appears in the top 15\%, while white squares indicate that proposals were submitted but none ranked within this subset. Red squares denote that no proposals with that keyword were submitted in that cycle.

In both panel and DPR cycles, not all keywords are represented among the top-ranked proposals. Furthermore, certain keywords are frequently absent across multiple cycles. This occurs most prominently in Categories~4 and~5, where some keywords attract relatively few submitted proposals.

The top panel of Figure~\ref{fig:diversity} shows the fraction of keywords that are present among the top 15\% of proposals by cycle. Keywords for which no proposals were submitted in a given cycle are excluded, so the percentages reflect the fraction of \emph{submitted} keywords represented among the top 15\% of proposals. Averaged over all cycles, approximately 87\% of keywords are represented in the top-ranked group, with no systematic trends related to the cycle or the review process: the mean fraction is 87\% for both panel reviews and DPR.

While Figures~\ref{fig:diversity_matrix} and~\ref{fig:diversity} quantify scientific diversity in terms of topical breadth, we separately assess whether individual keywords are represented among the top-ranked proposals at different rates under panel review and DPR. As shown in Appendix~\ref{app:keyword_top15}, a small number of keywords exhibit statistically significant differences. However, these isolated effects show no coherent pattern across scientific 
categories, and we therefore find no evidence for a systematic shift in scientific emphasis between the two review systems.

\subsection{Analysis Using Topic Modeling}
While keywords provide a standardized classification scheme, they reflect selection choices made by proposers and may not fully capture the nuanced scientific content of a proposal. As a complementary approach, we analyze proposal diversity using unsupervised topic modeling, which provides a data-driven representation of the scientific topics present in ALMA proposals.

We apply LDA to the titles, abstracts, and scientific justifications of all submitted proposals, following the methodology developed for the ALMA proposal assignment system \citep{Carpenter25}. The number of latent topics is fixed to 100 to provide a flexible and granular representation of the proposal corpus across all cycles. We verified that varying the number of topics between 50 and 500 does not introduce systematic differences in topic coverage between panel review and DPR, indicating that the comparison between review systems is robust to topic model granularity.

The bottom panel of Figure~\ref{fig:diversity} shows the fraction of these 100 topics represented among the top 15\% of proposals in each cycle. Consistent with the keyword-based analysis, the fraction of topics remains approximately constant across cycles and shows no systematic dependence on the review process. Averaged over all cycles, approximately 77\% of topics are represented among the top-ranked proposals, with similar results for panel reviews (75\%) and DPR (79\%). 

\section{Ranking Dispersion and Reviewer Agreement}
\label{sec:ranks}

The transition to DPR raises questions regarding the consistency of proposal evaluations in the absence of a collective panel discussion. Because DPR relies on independent assessments from a larger and more heterogeneous reviewer pool, it is important to quantify whether this process increases the dispersion of individual evaluations relative to traditional panels. Furthermore, while these individual assessments were historically internalized within the panel process, the DPR process provides PIs with all ten individual reviews and their associated ranks. This transparency makes the underlying dispersion much more visible to the community, making it essential to compare its magnitude with that seen in traditional panel-based reviews.

In this section, we characterize reviewer agreement and ranking dispersion in DPR and place these results in context using benchmarks derived from panel review behavior. We first introduce statistical measures to quantify dispersion among reviewers. We then compare the observed DPR results with idealized simulations representing limiting cases of reviewer behavior. Finally, we compare the dispersion measured in DPR with that found in both Stage~1 and Stage~2 of the panel review process to quantify the extent to which the DPR process matches the variance of individual panel assessments and to measure the reduction in dispersion achieved by the panel discussion phase. We conclude by examining pairwise agreement between reviewers as a model-independent complement to the simulation-based comparisons.

\subsection{Measuring Reviewer Agreement}
\label{subsec:agreement}

We characterize reviewer agreement using two complementary measures: (1) the root-mean-square (RMS) dispersion of the individual ranks assigned to a proposal, and (2) the rank spread, defined as the difference between the poorest and best individual ranks assigned by the reviewers. Both measures quantify the degree to which reviewers agree on the relative merit of a proposal, while remaining sensitive to different aspects of the rank distribution.

\begin{figure}
\centering
\includegraphics[width=\columnwidth]{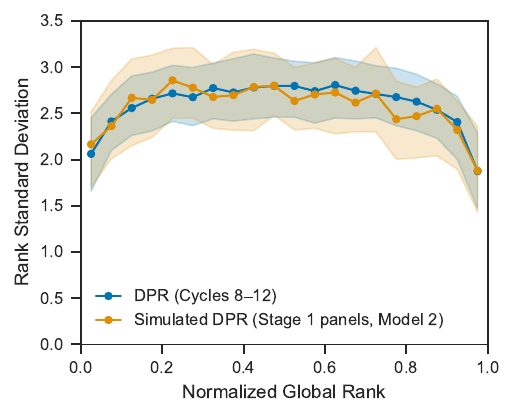}
\caption{
Median dispersion of individual reviewer ranks as a function of the aggregate proposal rank. Solid curves show the median RMS of the individual ranks assigned to each proposal, computed in bins containing 5\% of proposals ordered by aggregate rank. Shaded regions indicate the interquartile range (middle 50\%) of RMS values within each bin. Results from the Cycles~8--12 DPR process are shown in blue, while the orange curve shows simulations using Model~2 with Stage~1 panel data (see Section~\ref{subsec:stage1_dispersion}). Aggregate ranks are normalized within each cycle from 0 (best-ranked) to 1 (worst-ranked).
}
\label{fig:rms_ranks_stage1}
\end{figure}

Figure~\ref{fig:rms_ranks_stage1} shows the median RMS of the individual ranks assigned to proposals in Cycles~8--12 as a function of the overall (aggregate) proposal rank, where the overall rank has been normalized between 0 (best) and 1 (poorest) for each cycle. The shaded region indicates the interquartile range (IQR) of these RMS values, representing the middle 50\% of proposals within each bin. The median RMS dispersion is 2.63 overall, ranging from $\sim$1.9 for the bottom 5\% of the distribution to $\sim$2.8 in the middle, and decreasing to $\sim$2.1 for the best-ranked proposals, corresponding to typical reviewer disagreements of several rank positions within a ten-proposal set.

\begin{figure}
\centering
\includegraphics[width=\columnwidth]{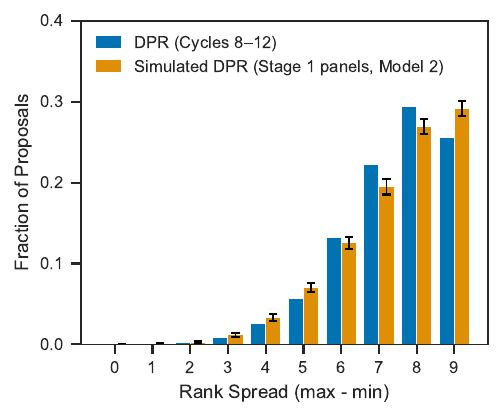}
\caption{
Distribution of rank spreads for individual proposals, comparing DPR and panel-based simulations. The spread is defined as the difference between the poorest and best individual ranks assigned by reviewers of a given proposal. The blue histogram shows results from Cycles~8--12 DPR, while the orange histogram shows results from 1000 simulations using Model~2 with Stage~1 panel data. A spread of zero indicates perfect agreement among reviewers, while the maximum possible spread of nine corresponds to proposals receiving both the best (1) and poorest (10) ranks. Error bars on the orange histogram represent the RMS across simulations.
}
\label{fig:spread_ranks_stage1}
\end{figure}

As an alternative metric, Figure~\ref{fig:spread_ranks_stage1} presents the distribution of rank spreads for proposals from Cycles~8--12. A rank spread of zero indicates that all reviewers assigned the identical individual rank to a proposal, while the maximum possible spread of nine corresponds to proposals that received both a rank of 1 and a rank of 10. Large rank spreads are remarkably common: 77\% of proposals exhibit a rank spread between 7 and 9, and approximately 26\% received both the best and worst possible ranks from different reviewers. This implies that for more than a quarter of proposals, reviewers completely disagreed on whether the proposal belonged at the top or bottom of their assigned set.

These results demonstrate that substantial dispersion in individual ranks is a pervasive feature of independently reviewed proposals under DPR. To determine whether the magnitude and structure of this dispersion are consistent with expectations, we compare these observed results with simulations representing both idealized scenarios and historical panel behavior.

\subsection{Comparison with Idealized Ranking Simulations}

To place the observed DPR rank distributions in context, we construct two idealized simulations representing limiting scenarios (see also \citealt{DonovanMeyer22}). These simulations establish qualitative benchmarks for interpreting the magnitude of ranking dispersion.

In the first scenario (the ``random" case), each reviewer assigns ranks 1 through 10 to their assigned proposal set completely at random. Because every proposal has an equal probability of receiving any rank, the Central Limit Theorem ensures that the distribution of \emph{average} ranks across the 10 reviewers is approximately Gaussian and clusters around the mean rank of 5.5. In the second scenario (the ``perfect agreement" case), we assume proposals possess an intrinsic scientific merit that is perfectly and identically perceived by all reviewers. In this limit, a proposal's rank is determined solely by its merit relative to the other nine proposals in its assigned set, resulting in a broad, nearly uniform distribution of average ranks across the pool.

\begin{figure*}
\centering
\includegraphics[width=\textwidth]{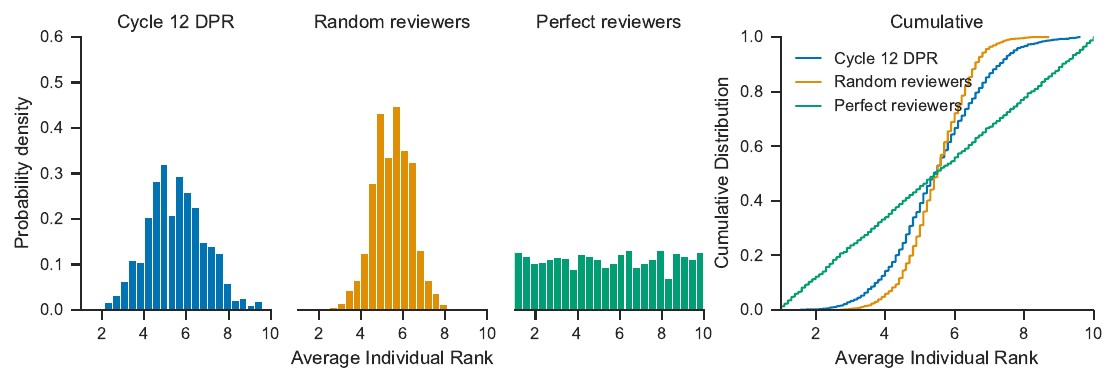}
\caption{
Comparison of average individual ranks for Cycle~12 DPR and two idealized scenarios. First three panels show probability distributions for: (1) actual Cycle~12 DPR results, (2) random reviewer assignment simulation, and (3) perfect reviewer agreement with intrinsic scientific merit. The rightmost panel shows cumulative distributions for all three cases, providing qualitative benchmarks for interpreting observed DPR rank distributions.
}
\label{fig:simulated_ranks}
\end{figure*}

Figure~\ref{fig:simulated_ranks} highlights the contrast between these two limits using Cycle~12 data. The actual Cycle~12 DPR distribution (first panel) is closer to the random case (second panel) than to the limit of perfect agreement (third panel). This proximity to the random baseline underscores the substantial dispersion inherent in independent reviewer assessments. 

However, as illustrated by the cumulative distributions in the rightmost panel, the DPR results are clearly distinct from the random simulation. The DPR distribution is broader, with an excess of proposals receiving consistently best-ranked and poorer-ranked average scores. This indicates that DPR is identifying common signals of proposal quality that would be absent in a purely random system.

\subsection{Comparison with Ranks Simulated from Stage~1 Panel Scores}
\label{subsec:stage1_dispersion}

Figure~\ref{fig:simulated_ranks} demonstrates that DPR identifies meaningful differences in aggregate proposal rankings, but with substantial reviewer-to-reviewer variance falling between the random and perfect agreement limits. To determine whether this level of dispersion is typical for independent scientific assessments, we compare the DPR results with historical panel reviewer behavior.

A direct comparison is complicated by the different mechanics of the two systems: traditional panel reviewers assign numerical scores to a large pool of proposals ($\sim$100), whereas DPR reviewers assign ranks to sets of ten. To bridge this gap, we first use historical Stage~1 (pre-discussion) scores to generate simulated rank assignments that replicate the DPR process using panelist evaluations. This methodology, detailed in Appendix~\ref{app:sim_panel_ranks}, constructs three simulation models (Models~1--3) that differ in their treatment of proposal-specific agreement. The simulations are not tuned to reproduce the DPR results; they provide an independent benchmark based solely on historical panel scoring behavior. Model~2, which accounts for the fact that some proposals naturally elicit stronger consensus among reviewers than others, provides the most realistic comparison and is used in the figures presented here.

The simulated panel-based distributions (orange curves and histograms in Figures~\ref{fig:rms_ranks_stage1} and \ref{fig:spread_ranks_stage1}) closely resemble the observed DPR results (blue). Quantitatively, the median RMS dispersion over all proposals is 2.63 for DPR and 2.59 for the Stage~1 panel simulations, a difference of 1.5\%. This similarity is not a mechanical artifact of the ranking framework: as shown in Figure~\ref{fig:simulated_ranks}, perfect reviewer agreement would produce a qualitatively different, nearly uniform distribution of average ranks. The close quantitative agreement between the DPR results and the panel-based simulations demonstrates that the level of dispersion observed under DPR is consistent with what panel reviewers themselves exhibited at Stage~1, prior to any discussion.

\subsection{Effect of Panel Discussion on Rank Dispersion: Stage~1 versus Stage~2}
\label{subsec:stage2_dispersion}

Panel-based reviews differ from the DPR framework in the nature of their second stage. While both processes allow for rank adjustments following the initial assessment, the mechanisms are distinct: DPR Stage~2 involves independent ranking of a broader set of proposals, whereas traditional panel Stage~2 involves in-person discussion and is designed to produce a consensus ranking. Importantly, producing a consensus does not require that all disagreements are eliminated; reviewers may still hold differing views on the relative merit of proposals, even though a single ranked outcome is assigned.

In practice, Stage~2 DPR reviews result in relatively few changes. Across Cycles~8--12, only $\sim$6.5\% of proposals had their rankings adjusted between Stage~1 and Stage~2, leaving the overall dispersion largely unchanged (see Appendix~\ref{app:stage12_comparison_dpr}). In contrast, panel Stage~2 deliberation explicitly considers the full set of untriaged proposals and is intended to align reviewer perspectives, typically reducing rank dispersion relative to Stage~1, though substantial differences of opinion often remain.

\begin{figure}
\centering
\includegraphics[width=\columnwidth]{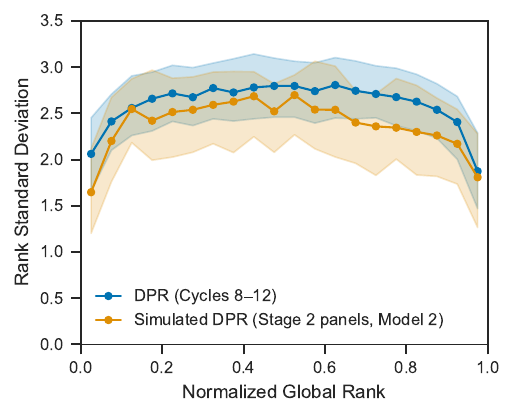}
\caption{
Similar to Figure~\ref{fig:rms_ranks_stage1}, but using Model~2 with Stage~2 panel data. The orange curve represents simulations based on Stage~2 (post-discussion) panel scores, reflecting the state of reviewer agreement after face-to-face deliberation. The dispersion observed in the DPR process is approximately 8.7\% higher than that achieved through traditional consensus-building across most of the ranking distribution.
}
\label{fig:rms_ranks_stage2}
\end{figure}

Figure~\ref{fig:rms_ranks_stage2} compares the median RMS dispersion of Cycles~8--12 DPR results with Model~2 simulations based on Stage~2 (post-discussion) panel scores. While the qualitative dependence of dispersion on aggregate rank is similar between the two datasets, the Stage~2 panel simulation exhibits a consistently lower level of dispersion across the entire ranking range. Quantitatively, the median RMS dispersion over all proposals is 2.63 for DPR and 2.40 for the Stage~2 panel simulations, a difference of 8.7\%. 

\begin{figure}
\centering
\includegraphics[width=\columnwidth]{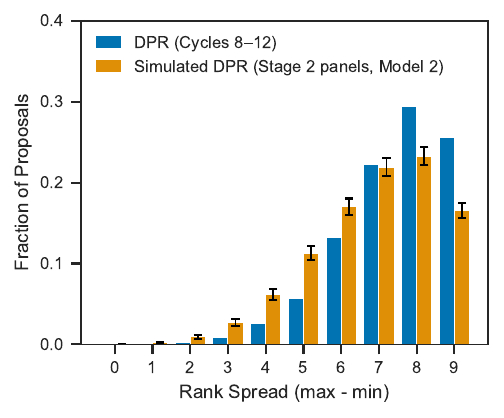}
\caption{
Similar to Figure~\ref{fig:spread_ranks_stage1}, but using Model~2 with Stage~2 panel data as input to the simulated DPR; the actual DPR rankings (blue) remain the same.
}
\label{fig:spread_ranks_stage2}
\end{figure}

Figure~\ref{fig:spread_ranks_stage2} compares the distribution of rank spreads observed under DPR (Cycles~8--12) with a DPR-like distribution simulated from Stage~2 (post-discussion) panel scores in Cycles~0--7. As expected, the simulated Stage~2 distribution is shifted toward smaller rank spreads, reflecting the increased alignment among panelists following discussion. Even after discussion, however, a considerable level of dispersion persists: 17\% of proposals still exhibit the maximum rank spread of 9, and 62\% maintain rank spreads between 7 and 9. This indicates that while panel discussion can align reviewer perspectives to some extent, substantial differences of opinion often remain. 

\subsection{Pairwise Agreement Between Reviewers}

While rank dispersion characterizes the spread of individual evaluations assigned to a proposal, it does not directly probe whether reviewers agree on the relative ordering of proposals. We therefore consider a complementary metric based on pairwise proposal orderings, which measures how often two reviewers assign the same relative ranking to proposals they have both evaluated. This approach compares reviewer judgments directly, without requiring panel modeling or assumptions about proposal population differences, providing a model-independent complement to the simulation-based comparison presented in Section~\ref{subsec:stage1_dispersion}.

For a given set of proposals, we consider all unordered pairs $(i,j)$ that are evaluated by the same two reviewers. For each proposal pair, we assign an agreement indicator equal to 1 if both reviewers assign the same relative ordering to the pair, and 0 if they assign opposite orderings. The pairwise agreement fraction is then defined as the mean of this indicator over all evaluated proposal pairs, with $P_{\mathrm{agree}} = 1$ corresponding to perfect agreement and $P_{\mathrm{agree}} = 0$ to systematic reversal of relative ordering. Under this convention, a purely random ranking produces an expected agreement fraction of $P_{\mathrm{agree}} = 0.5$. Uncertainties on the pairwise agreement fractions are quoted as 68\% confidence intervals, computed assuming binomial statistics for the number of agreeing proposal pairs and treating each pairwise comparison as an independent sample.

In DPR, reviewers provide strict rank orderings and ties do not occur. In panel-based reviews, reviewers assign numerical scores rather than explicit ranks; in this case, we assume that score ordering reflects reviewer priorities. Proposal pairs receiving identical scores from one or both reviewers are assigned an agreement indicator of 0.5, corresponding to the expectation value for random ordering.

\begin{figure}
\centering
\includegraphics[width=\columnwidth]{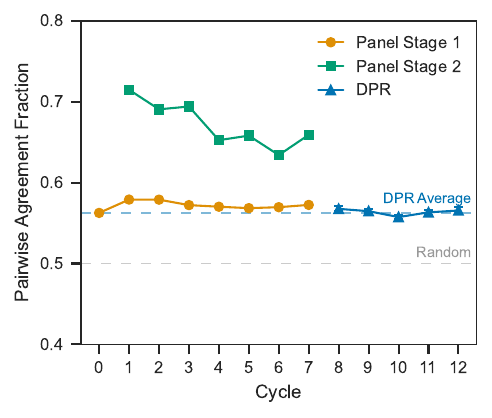}
\caption{
Pairwise agreement fraction as a function of cycle for Stage~1 panel reviews, Stage~2 panel reviews, and DPR. The agreement fraction measures the probability that two reviewers assign the same relative ordering to pairs of proposals they have both evaluated. The horizontal dashed line indicates the expected agreement for random rankings. Agreement levels for all review modes are significantly above random expectation. The agreement fractions observed for DPR are comparable to those obtained from Stage-1 panel-based reviews.
}
\label{fig:pairwise_by_cycle}
\end{figure}

Figure~\ref{fig:pairwise_by_cycle} shows the evolution of the pairwise agreement fraction as a function of cycle for Stage~1 panel reviews, Stage~2 panel reviews, and DPR. For all review modes, agreement fractions are significantly above the expectation for random rankings ($P_{\mathrm{agree}} = 0.5$) and exhibit no abrupt changes with cycle. The agreement levels observed for DPR are closely matched to those measured for Stage~1 panel reviews, with typical agreement fractions in the range $P_{\mathrm{agree}} \simeq 0.56$--0.57 across all cycles. Stage~2 panel reviews exhibit higher agreement fractions ($P_{\mathrm{agree}} \simeq 0.63$--0.71), reflecting the increased convergence introduced by panel discussion and re-scoring.

To examine whether reviewer agreement within DPR depends on reviewer career stage, we compute the pairwise agreement fraction separately for each pairing of reviewer career status. Career stage is determined from the year in which each reviewer obtained their PhD, based on information provided in user profiles and records online. The career status was obtained for 98\% of the reviewers. The career-dependent agreement analysis presented here is restricted to proposal pairs for which career status information is available for both reviewers.

\begin{figure}
\centering
\includegraphics[width=\columnwidth]{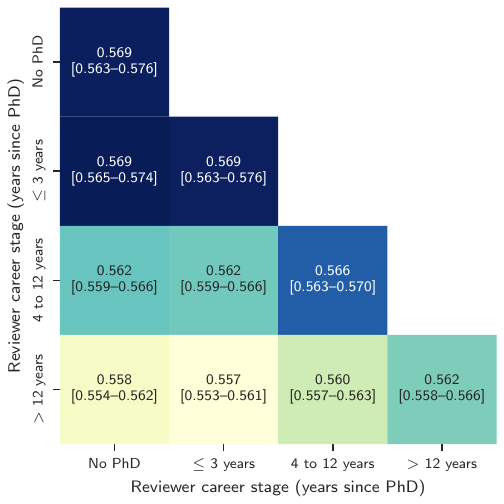}
\caption{
Pairwise agreement fraction within DPR as a function of reviewer career status. Each cell shows the agreement fraction for a given pairing of career levels, with 68\% confidence intervals indicated in brackets. Agreement fractions span a narrow range across all career-level combinations. For most career groups, agreement is marginally highest within the same career level, while the lowest agreement occurs between the most widely separated career stages. Overall differences are small in magnitude.
}
\label{fig:career_matrix}
\end{figure}

Figure~\ref{fig:career_matrix} shows the resulting agreement matrix for the DPR cycles. Agreement fractions span a narrow range across all career-level combinations (0.557--0.569). For most career groups, agreement is marginally higher within the same career level, while the lowest agreement is observed between the most widely separated career stages (i.e., students and the most senior reviewers). Although some of these differences are statistically detectable given the large number of proposal pairs, their magnitude is small (less than 1\%), indicating that reviewer agreement in DPR is largely independent of career stage.

\section{Assessment of Review Comment Quality}
\label{sec:quality}

Beyond proposal rankings, the effectiveness of any peer-review system depends on the quality of the written feedback provided to proposers. Written comments play a central role in communicating scientific strengths and weaknesses of a given proposal, guiding future resubmissions, and promoting transparency and accountability. The transition from panel-based review to DPR therefore raises important questions about whether the nature and perceived quality of reviewer feedback have changed.

In panel reviews, a single consensus report is provided to PIs, written by the primary reviewer. For proposals that proceed to Stage~2, this report reflects the outcome of collective panel discussion; for triaged proposals, it is based on the Stage~1 individual assessments alone. In contrast, DPR relies entirely on independently generated comments from individual reviewers, which are sent verbatim to the PIs. This structural difference has the potential to affect both the content and variability of review feedback.

In Cycles~8--10, the first three cycles using DPR, PIs were asked to evaluate the quality of the written comments they received. Those surveys indicated that 22\% of individual comments were rated as \emph{very helpful}, 49\% as \emph{somewhat helpful}, 27\% as \emph{inaccurate or unhelpful}, and 2\% as \emph{unprofessional}. However, these assessments were found to be correlated with review outcomes: comments associated with more favorable individual reviewer ranks were typically rated more positively than those associated with less favorable ranks \citep{DonovanMeyer22}. One likely contributor to this outcome-related bias is that PIs were shown both the written comment and the associated individual reviewer rank when completing the survey, so that perceptions of review quality may have been influenced by the presence of an unfavorable rank on their own proposal \citep[see][and references therein]{Goldberg25}.

To avoid this outcome-related bias, in Cycle~12 \emph{reviewers} were asked to evaluate ten randomly selected reviews written by other reviewers for proposals within their assigned set. To focus the assessment on the quality of the written feedback itself, reviewers were shown only the written comments; neither the associated individual reviewer ranks, the identity of the review author, nor the career stage of the author were provided. This approach ensured familiarity with the scientific context while avoiding biases related to knowledge of review outcomes or reviewer identity. Reviewers were asked to classify the assigned reviews as \emph{high quality} (``The review provides clear, specific, and constructive feedback that effectively identifies the proposal's strengths and weaknesses.''), \emph{adequate} (``The review offers some useful insights but lacks the detail, clarity, or specificity needed to be fully effective.''), or \emph{low quality} (``The review fails to provide meaningful feedback, contains significant errors, or adopts an unprofessional tone.''). The rating categories used in this reviewer-based survey differ from those employed in the earlier PI assessments, precluding direct quantitative comparison between the two sets of results.

In total, 228 reviewers participated in the survey (21\% of eligible reviewers), providing 3420 individual review evaluations. The participating sample is representative of the broader reviewer pool: the distributions of regional affiliation, gender, and career stage among survey participants are similar to those of all Cycle~12 reviewers. Because participation in the survey was voluntary, however, some degree of non-response bias cannot be excluded. Compared to non-participants, survey participants had a higher Stage~2 completion rate (92\% versus 55\%), wrote longer final review comments on average (987 versus 877 characters), and were more likely to revise at least one ranking (38\% versus 20\%) or comment (47\% versus 24\%) during Stage~2. These differences indicate that survey respondents were generally more engaged in the review process than non-respondents. Because the survey evaluated randomly assigned peer reviews rather than respondents' own reviews, the direction and magnitude of any resulting bias in the measured review-quality distributions are unclear.

\begin{figure}
\centering
\includegraphics[width=\columnwidth]{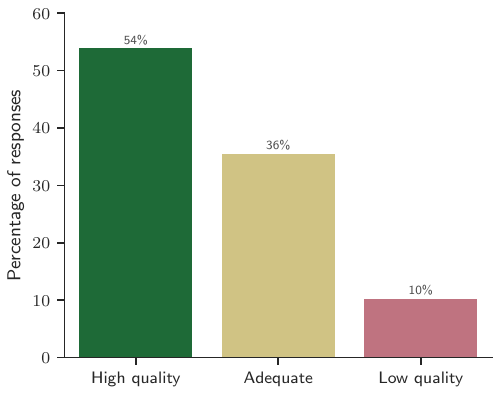}
\caption{Percentage of Cycle~12 reviews rated as high quality, adequate, or low quality by Cycle~12 reviewers based solely on the written comments.}
\label{fig:ratings_quality}
\end{figure}

Figure~\ref{fig:ratings_quality} summarizes the distribution of perceived review quality. Overall, 54\% of reviews were rated as high quality, 36\% as adequate, and 10\% as low quality. These results indicate that a majority of reviews are viewed by peers as constructive and informative, while also highlighting that a small fraction of reviews is viewed as low quality.

\subsection{Reviewer Career Stage}
\begin{figure}
\centering
\includegraphics[width=\columnwidth]{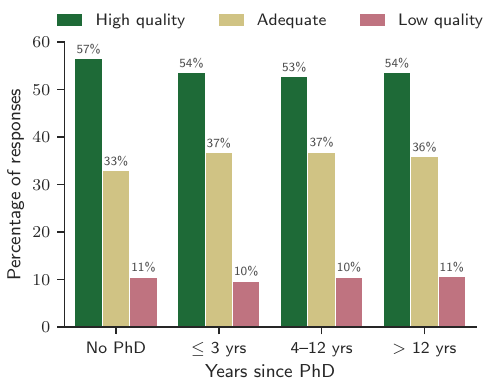}
\caption{Percentage of Cycle~12 reviews rated as high quality, adequate, or low quality as a function of the career stage of the reviewer who wrote the review. Ratings are based solely on the written comments.}
\label{fig:ratings_quality_writer}
\end{figure}

In ALMA’s implementation of distributed peer review, students and less-experienced postdoctoral researchers are permitted to serve as reviewers, a role they did not generally have in ALMA’s panel-based review process. Figure~\ref{fig:ratings_quality_writer} shows the perceived quality of reviews as a function of the career stage of the reviewer who wrote the comment. The distributions of high-quality, adequate, and low-quality reviews are similar across career stages, indicating that reviews written by students and postdoctoral researchers are, on average, rated comparably to those written by more senior researchers.

\subsection{Reviewer Expertise}
\begin{figure}
\centering
\includegraphics[width=\columnwidth]{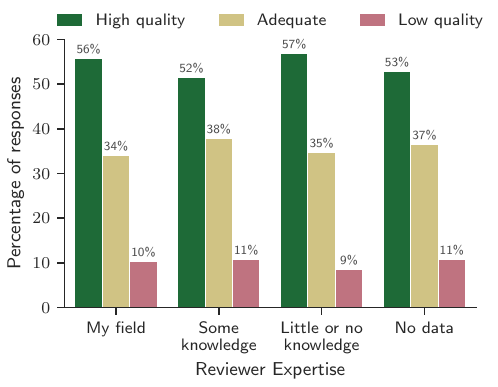}
\caption{Percentage of Cycle~12 reviews rated as high quality, adequate, or low quality as a function of the reviewer’s self-declared expertise for the proposal being reviewed. Ratings are based solely on the written comments.}
\label{fig:ratings_quality_expertise}
\end{figure}

Achieving good alignment between reviewer expertise and proposal topics is always a challenge in both panel-based review and DPR. Over the years in which the DPR process has been implemented, ALMA has monitored the level of expertise reviewers have relative to the proposals they were assigned to review by asking them to self-report their familiarity with each assigned proposal as ``my field,'' ``some knowledge,'' or ``little or no knowledge.'' Figure~\ref{fig:ratings_quality_expertise} shows the perceived quality of reviews as a function of the reviewer’s self-declared expertise. The results show little dependence of perceived review quality on expertise level. Reviews written by reviewers who report limited familiarity with the proposal topic are, on average, rated similarly to those written by reviewers who consider the topic to be within their primary field.

\section{Discussion}
\label{sec:discussion}

This study provides an empirical comparison of DPR and traditional panel-based review using data spanning 13 ALMA proposal cycles and more than 20,000 proposals. Our analysis focuses on aggregate statistical properties of proposal rankings, their dispersion, scientific breadth, and the quality of written reviews. In this section, we interpret the principal findings, discuss their limitations, and consider their implications for the design and evaluation of peer review systems in astronomy.

\subsection{Interpreting Differences in Ranking Systematics}

Section~\ref{sec:systematics} examined proposal ranking distributions under DPR and panel-based review across PI demographics, including region, gender, and experience, as well as technical characteristics such as observing mode, receiver band, and requested observing time. Most systematic patterns observed under panel review persist under DPR, indicating substantial continuity between the two systems. Where differences do appear, they are limited in scope and require careful interpretation. We find no evidence that the implementation of DPR introduces new systematic dependencies in proposal rankings beyond those already present under panel-based review. While the absence of detected effects does not preclude subtler systematics below the sensitivity of this dataset, the observed patterns are more consistent with pre-existing features of the review process than with effects specific to DPR.

\subsubsection{Regional Patterns}

Proposals submitted by East Asian PIs show a statistically significant improvement in rankings under DPR relative to panel-based review. As discussed in Section~\ref{subsec:region}, this improvement is absent in Cycle~8, first appears in Cycle~9, and strengthens through Cycle~12. This delayed onset and progressive nature provide important constraints on potential explanations.

Factors introduced simultaneously with DPR in Cycle~8, including the distributed review framework itself and the adoption of dual-anonymous review, cannot alone explain a trend that emerged in Cycle~9. Similarly, the introduction of machine-learning-based proposal assignment algorithms in Cycle~10 \citep{Carpenter25} postdates the initial improvement, though these algorithms may contribute to the continued trend in later cycles.

One possible mechanism is the mandatory review participation requirement inherent to DPR. Unlike panel-based review, where only a small subset of PIs served as reviewers, DPR requires a reviewer from every proposal team (usually the PI) to evaluate ten proposals from across the community. This exposure may provide insight into what constitutes a competitive proposal, which could then be applied to subsequent submissions. Such learning effects would emerge only after the first DPR cycle, consistent with improvements beginning in Cycle~9.

The increasing availability of language assistance tools may also play a role. ALMA's official policy permits PIs to use generative artificial intelligence (GAI) to assist with proposal preparation, while remaining fully responsible for the content. However, widespread GAI tools became available only around Cycle~10, well after the improvement began in Cycle~9, suggesting this cannot be the primary driver, though it may contribute to continued improvements in later cycles. Dual-anonymous review, introduced in Cycle~8, may have time-dependent effects if it mitigates unconscious biases that could have operated differently within regional allocation pools where proposals compete primarily for limited regional time.

Comparable improvements are not observed for Chilean PIs or PIs from Other regions, despite these groups being subject to the same changes in review format, anonymity, and assignment algorithms. The regional specificity of the improvement is not fully explained by any of the mechanisms discussed above, since these mechanisms apply equally to all regional groups.

\subsubsection{PI Experience}

The most experienced PIs, defined here as those submitting proposals in every available cycle, receive significantly poorer rankings under DPR than under panel-based review. Previous work \citep{Carpenter22} showed that the relative rankings of experienced PIs declined following the introduction of partial anonymization in Cycle~7, before DPR was implemented. The persistence of this effect through the DPR era is therefore more consistent with the introduction of anonymization beginning in Cycle~7 than with the change in review structure itself.

\subsubsection{Requested Observing Time}
\label{subsec:discussion:obstime}

Systematic differences are observed as a function of requested observing time. Proposals requesting 40--50~h receive significantly higher rankings under DPR than under panel-based review, while proposals requesting less than 10~h receive poorer rankings under DPR than under panels. In the early ALMA cycles, the observatory explicitly encouraged short proposals in order to maximize community access and mitigate technical and operational risk. As ALMA matured, the expanding body of published results enabled increasingly sophisticated science programs, and by Cycle~8 the observatory actively encouraged more ambitious proposals requiring longer integration times. The observed shift toward more favorable rankings for large time requests under DPR is therefore more consistent with evolving proposal guidance than with an effect of the review process itself.

\subsubsection{Other Technical Characteristics}

A small number of additional technical characteristics show differences between panel-based review and DPR. Proposals requesting full polarization and those requesting Band~8 observations receive lower rankings under DPR than under panel-based review. For Band~8 proposals, this difference is driven primarily by the first two cycles it was offered (see Section~\ref{subsec:band}), suggesting a transient effect associated with the initial availability of this capability rather than a persistent feature of the review system. However, the absence of similar patterns in other newly introduced capabilities suggests this is not a general feature of capability rollout.

The cycle-by-cycle dependence of the full-polarization effect (see Section~\ref{subsec:modes}) indicates that the aggregate panel--DPR difference does not reflect a simple change in review process. Full-polarization proposals show the highest rankings during the single-anonymous panel era (Cycles~2--6), a reduced effect in Cycle~7 when investigator names were first randomized prior to the introduction of DPR, and no consistent signal across DPR cycles. The decline of the effect coinciding with the partial anonymization in Cycle~7 suggests that name visibility may have contributed to the early-cycle advantage. However, the absence of comparable behavior in other specialized observing modes argues against name visibility as a sufficient explanation on its own. This pattern suggests that the early-cycle enhancement is more consistent with factors that vary across cycles rather than by a persistent difference between review systems, and the origin of the early-cycle advantage remains uncertain. A multivariate analysis controlling for proposer demographics and experience could in principle help isolate mode-specific effects, but the close coupling between procedural changes and the transition to DPR may limit the ability to disentangle these effects cleanly.

\subsection{Scientific Diversity and the Role of Discussion}
\label{subsec:discussion:diversity}

Section~\ref{sec:diversity} finds no measurable difference in the scientific diversity of top-ranked proposals between panel-based review and DPR, based on both proposer-assigned keywords and unsupervised topic modeling. This result may seem counterintuitive because panel discussions allow reviewers to consider scientific balance across a large number of proposals, whereas DPR reviewers evaluate a small set of proposals independently without seeing the broader proposal landscape. The structure of the ALMA review process may help explain why this does not lead to measurable differences, as panel-level normalization is applied to maintain representation across all science categories.

The similarity of diversity metrics between panels and DPR suggests that combining a sufficient number of independent assessments from the same scientific community is enough to recover community-wide priorities, even without discussion-based balancing. Notably, this holds regardless of whether category-level normalization was applied in DPR (introduced in Cycles~10--12 but absent in Cycles~8--9). Therefore, the degree to which panels actively enforce diversity within their panel does not strongly affect the diversity of final rankings.

This does not mean panel discussions have no value. The metrics used here capture only coarse-grained topical representation and do not address more nuanced dimensions such as methodological approaches, conceptual novelty, or support for emerging and high-risk subfields. Panel discussions may influence these aspects in ways our analysis cannot detect.

\subsection{Rank Dispersion as a Feature of Independent Assessment}

Section~\ref{sec:ranks} shows that the dispersion of individual rankings observed under DPR is comparable to that found in Stage~1 panel simulations prior to discussion. This similarity suggests that the substantial rank dispersion observed under DPR is consistent with the inherent variability of independent scientific assessment rather than a consequence of the distributed review process itself. This pattern aligns with previously documented variability in scientific assessments in astronomy \citep{Patat18} and in other fields \citep{Mayo06,Pier18,Jerrim23}.

Panel deliberation reduces dispersion relative to independent assessment: the median RMS dispersion decreases from 2.63 under DPR to 2.40 in Stage~2 panel simulations. However, the reduction is incomplete, indicating that panel discussion narrows but does not eliminate variation in reviewer assessments. Even after discussion, a proposal placed in the top decile by one reviewer retains an 18\% probability of being placed in the lower half by another (see Figure~\ref{fig:agreement_matrix} in Appendix~\ref{app:sim_panel_ranks}). This reduction in within-panel variance is distinct from inter-panel consistency. Independent studies in which the same proposals are evaluated by separate expert panels have shown that final outcomes can differ substantially between panels, even when each achieves internal convergence \citep[e.g.,][]{Cole81,Obrecht07,Pier18}. Panel discussion reduces within-panel variance but does not eliminate dependence on panel composition and deliberative dynamics.

This distinction has direct consequences for how dispersion is experienced by proposers. In the ALMA implementation of DPR, PIs receive the individual rankings and written comments from assigned reviewers together with a global percentile rank, whereas in panel-based review they receive a summary of strengths and weaknesses and a final quartile, with individual scores undisclosed. ALMA's use of forced ranking in DPR further requires reviewers to place proposals explicitly within the full dynamic range of the assignment set, whereas individual panel scores do not, on their own, convey the same relative context. This transparency has both advantages and disadvantages: it allows PIs to distinguish consensus concerns from isolated opinions, but it also exposes them to outlier views that panel discussion might otherwise have moderated \citep{Oxley25}.

\subsection{Review Quality}

The Cycle~12 assessment indicates that peer reviewers rated 54\% of DPR reviews as high quality, 36\% as adequate, and 10\% as low quality. These ratings reflect subjective evaluations by fellow reviewers based solely on the written comments, rather than an external or objective standard of review quality. The analysis further shows that perceived review quality does not vary systematically with reviewer career stage, indicating that any career-stage dependence is small relative to the sensitivity of the present dataset.

A direct, like-for-like comparison with the quality of ALMA panel consensus reports is not available, but prior surveys suggest no strong evidence of systematic differences in perceived review quality between DPR and panel-based systems. Following the Cycle~7 Supplemental Call, 31\% of PIs rated DPR comments as better than panel reports, while 35\% rated them as worse and 25\% rated them as similar (with the remaining 9\% being first-time submitters who could not make the comparison); 157 PIs responded to the survey \citep{Carpenter20b}. ESO surveys likewise indicate that DPR reviews are often perceived as more helpful on average than consensus reports, with only $\sim$12\% rated as ``not helpful'' \citep{Jerabkova25}.

The fraction of reviews perceived as low quality under ALMA DPR also does not appear anomalously high when viewed in a broader context. Surveys of peer review in computer science conferences routinely find that 15--30\% of reviews are rated as unhelpful, with author perceptions strongly confounded by review positivity \citep[e.g.,][]{Prechelt18,Frachtenberg20}. In this light, the $\sim$10\% of low-quality reviews identified by reviewers in Cycle~12 falls within the range observed in other large-scale peer-review systems.

However, a key structural difference between review systems concerns how low-quality reviews are experienced by proposers. In panel-based review, each proposal yields a single consensus report; if 10\% of reports are low quality, only 10\% of PIs experience a poor review. Under DPR, each proposal receives ten independent reviews. Even if the underlying fraction of low-quality reviews remains at 10\%, the probability that a proposer receives at least one low-quality review in their package of ten is approximately 65\% ($=1-0.9^{10}$). This does not imply lower overall review quality, but rather reflects the increased visibility of reviewer heterogeneity.

At the same time, the absence of a panel discussion layer introduces structural incentives that can give rise to genuinely low-effort reviews, such as duplicating feedback across proposals, providing superficial commentary, or merely summarizing the proposal without critically evaluating its strengths and weaknesses. In the panel system, chairs and fellow reviewers provided collective oversight to curb such behavior. Under DPR, this responsibility falls primarily on the observatory, presenting a significant organizational challenge given the volume of reviews ($\sim$16,000 per cycle).

\subsection{Reviewer Workload and Community Participation}

In addition to its effects on ranking outcomes and perceived review quality, distributed peer review reshapes how reviewing effort is distributed across the scientific community. Panel-based review concentrates a substantial workload on a relatively small number of reviewers, who may be asked to evaluate on the order of $\sim$100 proposals per cycle and to commit to service across multiple cycles. DPR instead spreads the reviewing effort across a much larger pool, with each reviewer expected to evaluate a fixed number of proposals per cycle. While the per-cycle workload under DPR is smaller, the obligation persists for as long as investigators remain active proposers, representing a non-trivial and ongoing commitment. At the same time, this broader participation exposes a larger fraction of the community to the proposal evaluation process.

DPR therefore redistributes reviewer effort rather than eliminating it, trading episodic intensive service by a few for sustained moderate participation by many. This trade-off may carry both benefits and costs. The long-term consequences of this sustained obligation are not yet well understood, particularly in a highly oversubscribed environment where most proposals are declined. It remains an open question whether the expectation of continuous reviewing, coupled with low proposal success rates, affects community engagement, reviewer motivation, or perceptions of fairness over time. The mandatory nature of DPR participation may also contribute to the fraction of low-quality reviews identified in Section~\ref{sec:quality}, since reviewers who would otherwise decline participation may be less motivated to provide thorough feedback.

\subsection{Limitations of the Current Study}

This analysis is observational rather than experimental, and its conclusions must be interpreted accordingly. The transition from panel-based review to DPR coincided with several other substantive changes to the ALMA review process, including the adoption of dual-anonymous review, shifts in reviewer pool composition, changes in proposal--reviewer matching algorithms, and the natural evolution of scientific priorities within the community. As a result, observed differences between panel and DPR outcomes cannot be uniquely or cleanly attributed to the change in review structure alone.

Our results demonstrate that panel-based review and DPR yield statistically similar ranking distributions at the population level within the ALMA context. However, the absence of detected differences should not be interpreted as proof that the two systems are equivalent in all respects. Subtle effects below the sensitivity of the available data may exist but remain undetected, particularly for rare proposal types, small demographic subgroups, or outcomes that manifest only at the tails of the ranking distribution. The statistical power of the present analysis is necessarily limited in these regimes.

In addition, this study focuses on measurable properties of review outcomes (such as ranking distributions, dispersion, topical diversity, and perceived review quality) rather than on qualitative aspects of the review process itself. Panel-based review offers structured discussion, calibration, and synthesis, which may help identify errors, place outlier opinions in perspective, or refine collective judgments. However, it also faces scalability constraints and is susceptible to well-documented group dynamics and biases \citep{Lee13,Guthrie18,Oxley25}. DPR, by contrast, addresses scalability, distributes reviewer effort more broadly, and increases transparency, but it lacks an explicit discussion phase that might mitigate individual misunderstandings or extreme assessments. The present analysis does not attempt to adjudicate between these trade-offs.

Finally, this study evaluates the properties of the review process rather than its downstream scientific impact. The ultimate goal of any proposal review system is to enable high-impact scientific outcomes, but assessing the relationship between review structure and subsequent scientific productivity, citation impact, or legacy value lies beyond the scope of the data analyzed here. Such assessments would require long-term tracking of approved programs and careful control for confounding factors unrelated to the review process itself.

\section{Conclusions}
\label{sec:conclusions}

We present an empirical comparison of proposal review outcomes under ALMA's traditional panel-based system and its current distributed peer review framework, drawing on over 20,000 proposals and 160,000 individual reviews spanning Cycles~0--12. Because the transition to DPR coincided with the introduction of dual-anonymous review and broader community evolution, the findings reported here should be interpreted as descriptive rather than causal.

The systematic ranking patterns established under eight cycles of panel review are largely preserved under DPR. Proposals from North American and European PIs continue to receive above-average rankings, while those from Chilean, East Asian, and Other PIs tend to rank below average, with the notable trend of a progressive improvement for East Asian PIs beginning in Cycle~9. Proposals from first-time PIs continue to have the poorest ranks. The advantage that the most experienced PIs held under panel review is reduced under DPR in a manner consistent with the introduction of dual-anonymous review rather than the change in review structure itself. Technical characteristics such as observing mode, receiver band, and requested time show broadly similar ranking behavior across the two systems. Where differences appear, they are limited in scope and reflect trends that either predate distributed peer review, are confined to specific observing cycles, or may be influenced by contemporaneous proposal guidance, making it difficult to attribute them uniquely to the distributed format itself.

Scientific diversity among top-ranked proposals is similarly stable across the transition. Whether measured by proposer-assigned keywords or unsupervised topic modeling, the breadth of scientific topics represented in the top 15\% of proposals is comparable between the panel and DPR eras, with both systems achieving representation of approximately 87\% of submitted keywords. This indicates that aggregating independent assessments from a large, diverse reviewer pool reproduces community-wide scientific priorities without requiring a collective discussion phase.

The substantial rank dispersion visible to proposers under DPR is consistent with the inherent variability of independent scientific assessment. Simulations derived from Stage~1 panel scores reproduce the observed DPR dispersion closely, differing by only 1.5\%, and pairwise agreement between reviewers closely matches levels observed in Stage~1 panel assessments, demonstrating that comparable disagreement was already present in panel evaluations prior to discussion. Panel deliberation reduces this dispersion relative to Stage~1, but by only 8.7\% relative to DPR, and considerable variability persists even after discussion, consistent with documented heterogeneity in independent scientific assessment more broadly. DPR makes this inherent variability visible to proposers, by providing the full set of individual rankings and written comments rather than a single synthesized outcome.

Review quality assessments from Cycle~12 show that the majority of DPR reviews are rated as high quality or adequate by peers, with no dependence on the career stage or self-declared expertise of the reviewer who wrote the review. The 10\% of reviews rated as low quality nonetheless highlights the challenge of maintaining review quality standards across the $\sim$16,000 reviews produced each cycle. Under panel-based review, panel chairs and fellow panelists provided collective oversight that could identify and moderate low-quality reviews; no equivalent mechanism exists under DPR, placing this responsibility solely on the observatory.

Taken together, these findings indicate that DPR and panel-based review extract similar scientific signals from the community when viewed in aggregate. The two systems differ primarily in transparency, scalability, and how reviewing effort is distributed across the community, rather than in the ranking outcomes they produce. This consistency, sustained across five DPR cycles and a wide range of proposal characteristics, establishes a quantitative baseline for evaluating the distributed format and informs the ongoing design of proposal evaluation systems at ALMA and other large observatories.

\section*{Data Availability Statement}

The data underlying this article are not publicly available due to confidentiality restrictions associated with ALMA proposal review data. The derived data products underlying all figures in this paper, together with a utility script for reproducing the CDF figures, are publicly available at Zenodo \citep{Carpenter26data}. Methodological details sufficient to reproduce the statistical procedures are provided in the manuscript.

\software{Astropy \citep{astropy:2018}, SciPy \citep{Jones01}, \textsf{R} \citep{R}, kSamples \citep{Scholz19}}

\begin{acknowledgments}

We thank the anonymous referee for a careful reading of the manuscript and for constructive comments that improved the clarity, statistical rigor, and interpretation of the paper. We also thank Arnaud Belloche, Caitlin Casey, Mar\'ia D\'iaz Trigo, Jennifer Donovan Meyer, Bunyo Hatsukade, Adele Plunkett, Hideo Sagawa, Kazushi Sakamoto, and Catherine Vlahakis for their comments and suggestions. We are also grateful to the ALMA Science Advisory Committee for useful discussions about the ALMA peer review process.

\end{acknowledgments}

\bibliography{references}

\appendix
\section{Comparison of Stage~1 and Stage~2 DPR Ranks}
\label{app:stage12_comparison_dpr}

In DPR, reviewers may update their ranks and comments in Stage~2 after reading the anonymized comments from the other panel members. This appendix quantifies the impact of those revisions.

Across Cycles~8--12, 93.5\% of Stage~1 ranks were unchanged in Stage~2. Of the 6.5\% of the rankings that were modified, 69\% were changed by $\pm$ 1 rank and 18\% by $\pm2$ ranks. The impact on the highest-ranked proposals is similarly modest. On average, 6.3\% of proposals initially within the top 15\% overall move out of that tier after Stage~2 revisions. Thus, Stage~2 in DPR produces limited reshuffling even among the most competitive proposals.

\begin{figure}
\centering
\includegraphics[width=\columnwidth]{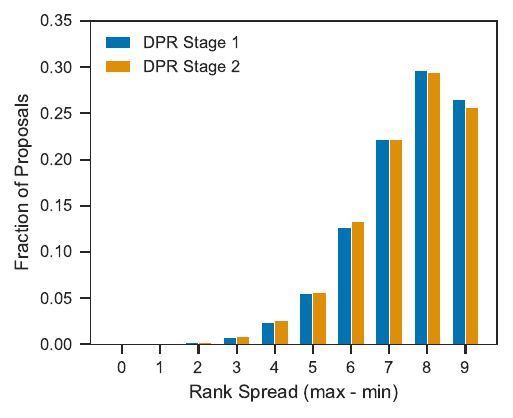}
\caption{
Distribution of proposal rank spreads in Stage~1 and Stage~2 of DPR for Cycles~8--12.}
\label{fig:stage12_dpr}
\end{figure}

Figure~\ref{fig:stage12_dpr} compares the distribution of rank spreads (see Section~\ref{subsec:agreement}) for Stage~1 and Stage~2. While there is a reduction in the number of proposals exhibiting the most extreme rank spreads, the overall distribution changes only slightly. These results indicate that Stage~2 revisions typically produce minor adjustments rather than substantial re-orderings.

\section{Comparison of Stage~1 and Stage~2 Panel Ranks}
\label{app:stage12_comparison_panels}

This appendix examines how proposal rankings change between Stage~1 (pre-discussion) and Stage~2 (post-discussion) of the panel review process, considering only proposals that were not triaged and therefore proceeded to Stage~2. Section~\ref{subsec:stage12_cdfs_panels} compares the cumulative rank distributions for each subgroup, assessing whether panel discussion systematically reshapes the population-level ranking patterns established during independent review. Section~\ref{subsec:stage12_spreads_panels} quantifies the effect of panel discussion on the relative ordering of individual proposals, focusing on the stability of the highest-ranked proposals and the reduction in extreme rank disagreements between Stage~1 and Stage~2.

\subsection{Cumulative Distribution Functions}
\label{subsec:stage12_cdfs_panels}

To enable comparison across cycles, ranks within each cycle have been normalized to the range [0,1] for Cycles~0--7. A note on interpreting Stage~2 rank distributions is warranted. Following Stage~1, the lowest-ranked proposals are removed through triage in a region-dependent fashion, and Stage~2 ranks are renormalized within the surviving proposal pool. As a result, statements about absolute rank levels within Stage~2 can be misleading; e.g., whether proposals from a given region rank above or below average. The focus here is therefore on whether panel discussion systematically reshapes the ranking patterns established in Stage~1, assessed by comparing the Stage~1 and Stage~2 distributions for each subgroup separately using the Anderson--Darling test.

\subsubsection{PI Regional Affiliation}
\label{subsec:stage12_region}

\begin{figure}
\centering
\includegraphics[width=\columnwidth]{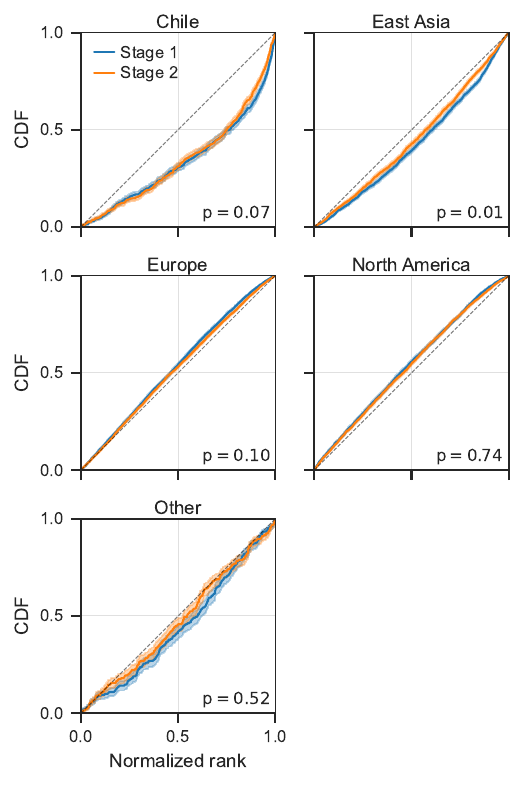}
\caption{Cumulative distribution functions of normalized proposal ranks for panel reviews in Cycles~0--7, shown separately for Stage~1 (pre-discussion) and Stage~2 (post-discussion) and grouped by PI regional affiliation. The $p$-value in each panel gives the result of a two-sample Anderson--Darling test comparing the Stage~1 and Stage~2 rank distributions for that region. The dashed line indicates a uniform distribution, corresponding to the absence of regional dependence in rankings.}
\label{fig:stage12_panels_region}
\end{figure}

Figure~\ref{fig:stage12_panels_region} compares the cumulative distributions of Stage~1 and Stage~2 ranks grouped by the regional affiliation of the proposal PI. For most regions, the Stage~1 and Stage~2 distributions are statistically indistinguishable, indicating that panel discussion does not systematically alter the regional ranking patterns established during independent review.

Notably, East Asian proposals show an improvement in their rank distribution between Stage~1 and Stage~2 ($p=0.01$), suggesting that Stage~2 deliberations are associated with an upward shift in the rankings of East Asian proposals relative to Stage~1. This effect is localized to a single regional group and does not alter the overall ordering of regional rank distributions, which remain broadly consistent between Stage~1 and Stage~2.

\subsubsection{PI Gender}
\label{subsec:stage12_gender}

\begin{figure}
\centering
\includegraphics[width=\columnwidth]{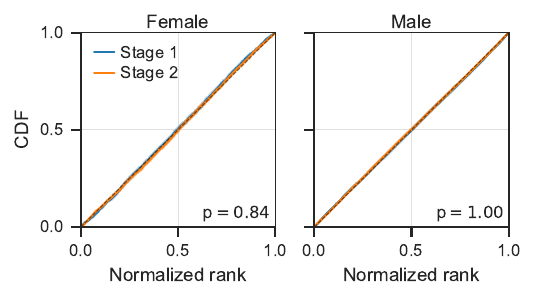}
\caption{Same as Figure~\ref{fig:stage12_panels_region}, but grouped by PI gender. No statistically significant differences are observed between the Stage~1 and Stage~2 rank distributions for any gender.}
\label{fig:stage12_panels_gender}
\end{figure}

Figure~\ref{fig:stage12_panels_gender} compares the cumulative distributions of Stage~1 and Stage~2 ranks grouped by PI gender. No statistically significant differences between Stage~1 and Stage~2 rank distributions are observed by gender, suggesting that discussion neither amplifies nor mitigates gender-related patterns established at Stage~1.

\subsubsection{PI Experience}
\label{subsec:stage12_experience}

\begin{figure}
\centering
\includegraphics[width=\columnwidth]{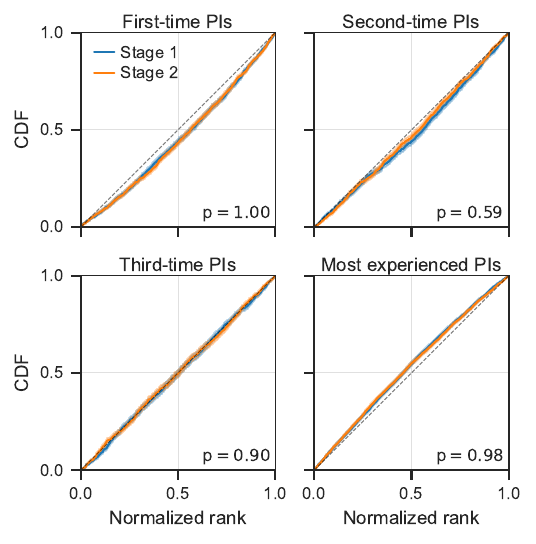}
\caption{Same as Figure~\ref{fig:stage12_panels_region}, but grouped by the PI’s prior experience submitting ALMA proposals (first time, second time, third time, or proposals submitted in every cycle). No statistically significant differences are observed between the Stage~1 and Stage~2 rank distributions for any experience group.}
\label{fig:stage12_panels_experience}
\end{figure}

Figure~\ref{fig:stage12_panels_experience} compares the cumulative distributions of Stage~1 and Stage~2 ranks grouped by the PI's prior experience with ALMA proposal submissions. The ranking patterns established in Stage~1 are largely preserved in Stage~2 for all experience groups. While second-time PIs show an improvement in their Stage~2 rankings compared to Stage~1, this difference is not statistically significant ($p=0.59$). Overall, these results indicate that Stage~2 deliberations largely preserve experience-based ranking patterns established during Stage~1, rather than systematically mitigating them, suggesting that panel discussion does not substantially compensate for experience-related differences established during independent review.

\subsubsection{Observing Modes and Special Categories}
\label{subsec:stage12_obsmode}

\begin{figure}
\centering
\includegraphics[width=\columnwidth]{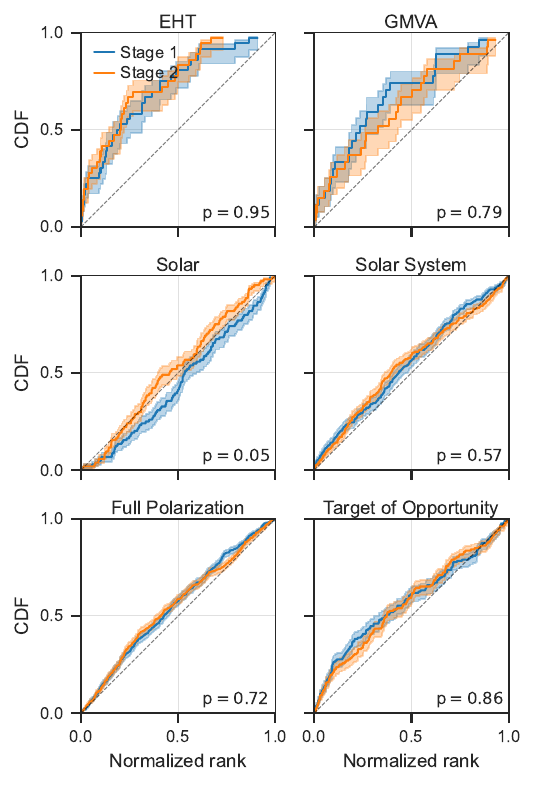}
\caption{Same as Figure~\ref{fig:stage12_panels_region}, but grouped by observing mode (EHT, GMVA, full-polarization), science area (Solar, Solar System), and proposal type (Target of Opportunity). No statistically significant differences are observed between Stage~1 and Stage~2 rankings.}
\label{fig:stage12_panels_obsmode}
\end{figure}

Figure~\ref{fig:stage12_panels_obsmode} compares the cumulative distributions of Stage~1 and Stage~2 ranks for proposals in special observing categories. These include special observing modes (EHT, GMVA, and full-polarization observations), specific science areas (Solar and Solar System observations), and specific proposal types (Target of Opportunity). No statistically significant differences are observed between Stage~1 and Stage~2 rankings. These results indicate that panel discussion does not systematically re-rank proposals in specialized observing categories, and that relative outcomes for these proposal types are largely established during the initial independent assessments.

\subsubsection{Receiver Band}
\label{subsec:stage12_rxband}

\begin{figure}
\centering
\includegraphics[width=\columnwidth]{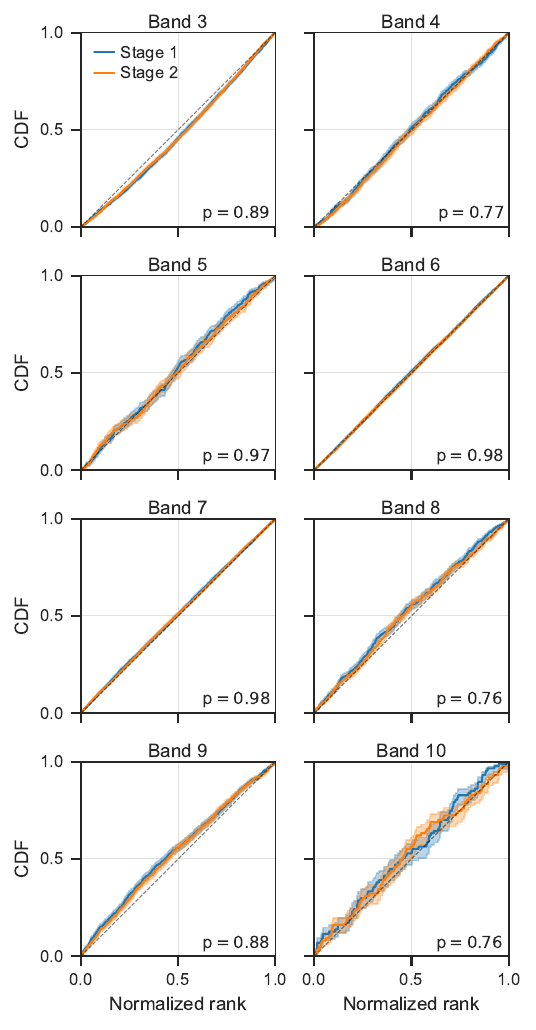}
\caption{Same as Figure~\ref{fig:stage12_panels_region}, but grouped by the primary receiver band requested. No statistically significant differences are observed between the Stage~1 and Stage~2 rank distributions for any band.}
\label{fig:stage12_panels_rxband}
\end{figure}

Figure~\ref{fig:stage12_panels_rxband} compares the cumulative distributions of Stage~1 and Stage~2 ranks grouped by the receiver band requested. Proposals requesting more than one receiver band will appear in multiple panels in this plot. No statistically significant differences between Stage~1 and Stage~2 rank distributions are observed for any receiver band, indicating that requested observing frequency does not become a stronger discriminator through panel discussion.

\subsubsection{Requested 12-m Array Time}
\label{subsec:stage12_obstime}

\begin{figure}
\centering
\includegraphics[width=\columnwidth]{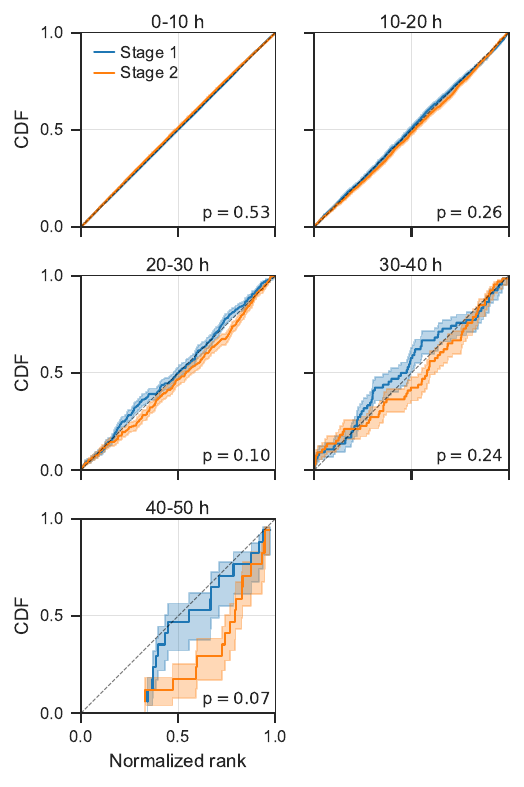}
\caption{Same as Figure~\ref{fig:stage12_panels_region}, but grouped by requested 12-m Array observing time.}
\label{fig:stage12_panels_obstime}
\end{figure}

Figure~\ref{fig:stage12_panels_obstime} compares the cumulative distributions of Stage~1 and Stage~2 ranks grouped by requested 12-m Array observing time. 
When proposals requesting 20--50~h are grouped together, Stage~2 rankings tend to be poorer than Stage~1 rankings ($p=0.02$), though this result does not survive correction for multiple comparisons within the time-request subsection. Nonetheless, this pattern is consistent with the fact that in the early panel cycles ALMA explicitly encouraged relatively small time requests, such that panel discussion may have reinforced community sentiment against larger programs rather than reflecting a systematic property of panel discussion itself.

\subsubsection{Summary}

These results demonstrate that the dominant systematic ranking patterns in panel review are established during independent assessment and persist through subsequent discussion. This finding justifies the use of Stage~1 panel rankings as a baseline for comparison with DPR outcomes in the main analysis, and reinforces the interpretation that many population-level features of proposal evaluation reflect the aggregation of independent expert judgments rather than the specific mechanics of panel discussion.

\subsection{Rank Spreads and Reordering}
\label{subsec:stage12_spreads_panels}

While the preceding subsections show that population-level ranking patterns are largely preserved between Stage~1 and Stage~2, panel discussion may still alter the relative ordering of individual proposals. To quantify this effect, we simulate DPR rankings based on panel scoring behavior following the procedure described in Section~\ref{subsec:stage1_dispersion}, considering only proposals that passed triage and were evaluated in both Stage~1 and Stage~2 (1264 proposals). Because this analysis is restricted to proposals that passed triage, the simulated Stage~1 rank spread distribution differs slightly from that shown in Figure~\ref{fig:spread_ranks_stage1}, which includes all Stage~1 proposals.

Based on Stage~1 panel scores, 30\% of proposals exhibit the maximum possible rank spread of 9. After Stage~2 panel discussion, this fraction decreases to 17\%, indicating that discussion reduces the most extreme disagreements among reviewers. A complementary measure of impact is the stability of the highest-ranked proposals: 31\% of proposals initially within the top 15\% based on Stage~1 scores would no longer occupy that tier after incorporating Stage~2 panel scores. This contrasts sharply with DPR (Appendix~\ref{app:stage12_comparison_dpr}), where only 6.3\% of top 15\% proposals change tier after Stage~2 revisions.

Thus, panel discussion narrows the most extreme rank spreads and produces greater reshuffling among the most competitive proposals than is observed in DPR. Nonetheless, considerable variation in relative assessments persists even after deliberation.

\section{Keyword-Level Comparisons Between Panel Review and DPR}
\label{app:keyword_top15}

As a supplementary diagnostic to the aggregate diversity measures presented in Section~\ref{sec:diversity}, we examine whether individual ALMA science keywords exhibit different rates of appearing among the top-ranked proposals under panel review and DPR.

For each of the 58 ALMA science keywords, we compute the percentage of submitted proposals ranked within the top 15\% under panel review (Cycles~1--7) and under DPR (Cycles~8--12). Uncertainties are estimated via bootstrap resampling over cycles. The uncertainty on the difference between review modes is computed as the quadrature sum of the two individual uncertainties. Statistical significance of differences between review modes is assessed using Fisher's exact test applied to the underlying proposal counts.

\begin{figure}
\centering
\includegraphics[width=0.96\columnwidth]{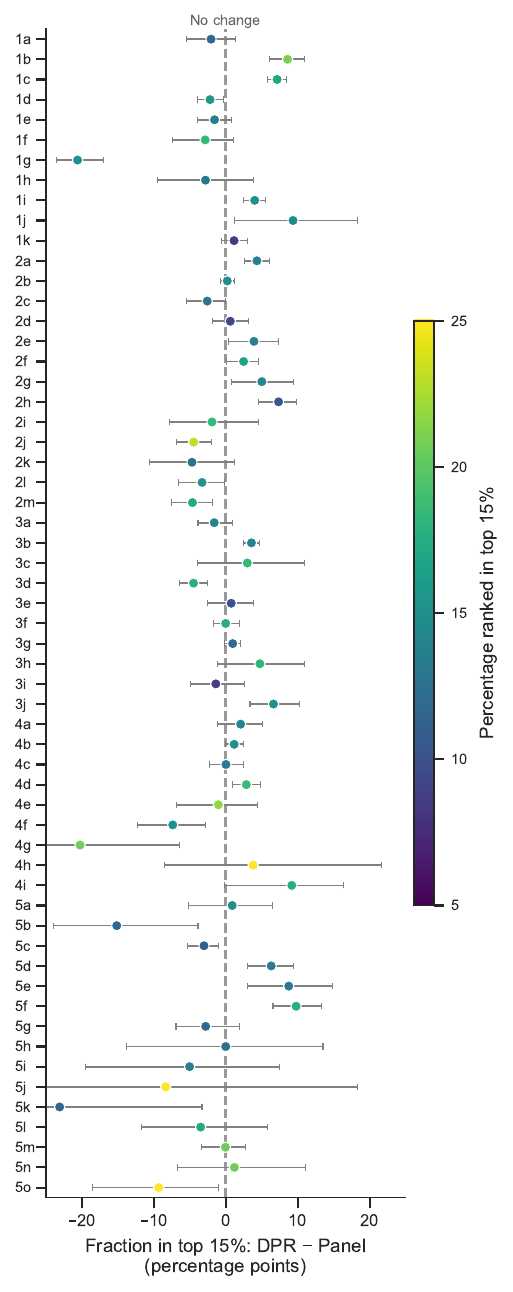}
\caption{Difference in the fraction of proposals ranked in the top 15\% between DPR and panel review (DPR~$-$~Panel), for each of the 58 ALMA science keywords. Error bars show 68\% bootstrap confidence intervals. Point colors indicate the average fraction of submitted proposals with a given keyword that ranked in the top 15\%, averaged across both review modes, as shown by the color bar. Three keywords (4g, 5j, 5k) have uncertainties extending beyond the axis limits. Keyword definitions are listed in \citet{Privon25}.}
\label{fig:kw_top15}
\end{figure}

Figure~\ref{fig:kw_top15} shows the difference in the top-15\% ranking percentage (DPR~$-$~Panel) for each keyword. The color of each point encodes how competitive proposals with that keyword are overall: keywords where a higher fraction of submissions reach the top 15\% (in both systems combined) appear in yellow, while those with lower overall success rates appear in blue. After correcting for multiple testing across 58 keywords using a false discovery rate of 1\%, no keywords exhibit statistically significant differences between panel review and DPR. At the single-test level, two keywords show differences with $p < 0.01$ (1c and 1g), and four additional keywords show marginal differences ($0.01 < p < 0.05$; 1b, 3b, 3d, and 5f), but these effects are not robust to multiple-testing correction and do not form a coherent pattern associated with either review mode.

Several keywords deviate from the nominal 15\% expectation in \emph{both} review systems, appearing consistently over- or under-represented among the top-ranked proposals regardless of review mode. These keywords fall near zero in the figure but show similarly high or low average top-15\% ranking percentages across both panel review and DPR (e.g., keywords 1k, 2d, and 3e are consistently under-represented, with average values of 6--10\%, while keywords 2j and 5o are consistently over-represented, with average values of 20--31\%). These persistent deviations reflect stable features of community evaluation patterns rather than changes introduced by the review process itself.

\section{Generating Proposal Ranks from Panel Reviewer Scoring Behavior}
\label{app:sim_panel_ranks}

This appendix describes the procedure used to simulate proposal rankings based on empirical panel reviewer scoring behavior. The purpose is to establish a statistical benchmark for the level of rank dispersion expected from independent reviewer assessments, given the degree of agreement observed among panel reviewers. By simulating what DPR rankings would look like if reviewers agreed at the same rate as panelists do before discussion, we can test whether the dispersion observed in DPR (Section~\ref{sec:ranks}) is unusual or simply reflects the inherent variability of independent scientific judgment. These simulations are not a fit to the DPR data. Rather, they provide an independent benchmark based on historical panel reviewer behavior. We evaluate three simulation models that differ in their assumptions about proposal-specific agreement levels. The details of each model and their construction are described in the subsections below.

In ALMA panel reviews, each reviewer assigns numerical scores to proposals on a scale from 1 (best) to 10 (poorest), with duplicate and fractional scores permitted. The central assumption of this analysis is that the relative ordering implied by these scores reflects each reviewer's internal ranking of proposals. We therefore characterize reviewer agreement in terms of relative ranking rather than absolute score values, and use this information to construct simulated proposal rankings for comparison to DPR.

\subsection{Reviewer Agreement Matrix}
\label{subsec:ram}

We quantify reviewer--reviewer agreement following the approach of \citet{Patat18}. For each reviewer $r$ and proposal $p$, we compute a quantile assignment based on the reviewer’s scores,
\begin{equation}
Q_r(p) = \mathrm{quantile}\!\left(\mathrm{rank}\big(s_r(p)\big)\right), \quad Q_r(p) \in \{1,2,\ldots,b\},
\end{equation}
where $b$ is the number of quantile bins used in the analysis. We use decile bins ($b=10$) to match the ranking scale used in DPR, but the results are not sensitive to the precise value of $b$.

We then count how frequently two reviewers assign the same proposal to given quantile bins. For each reviewer pair $(r_A, r_B)$ and proposal $p$ reviewed by both,
\begin{equation}
C_{ij} =
\sum_{(r_A, r_B)\in\mathcal{R}}
\sum_{p\in\mathcal{P}_{AB}}
\mathbf{1}\!\left(Q_{r_A}(p)=i \;\wedge\; Q_{r_B}(p)=j\right),
\end{equation}
where $\mathcal{P}_{AB}$ is the set of proposals reviewed by both reviewers.

\begin{figure*}
\centering
\includegraphics[width=\textwidth]{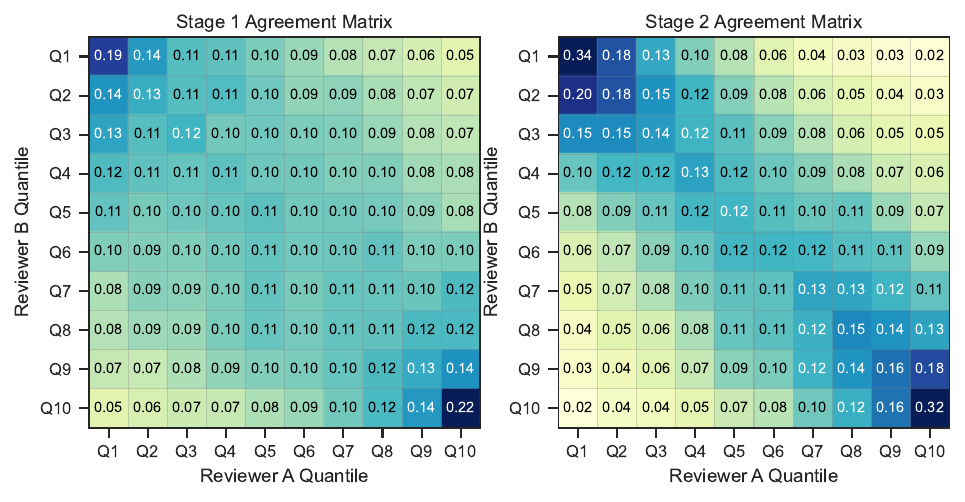}
\caption{Reviewer--reviewer agreement matrices derived from panel reviews in Cycles~0--7 for Stage~1 (left) and Stage~2 (right), using decile quantile bins.}
\label{fig:agreement_matrix}
\end{figure*}

The reviewer agreement matrix $M$ is defined as the conditional probability that Reviewer~B assigns a proposal to bin $j$ given that Reviewer~A assigns it to bin $i$. In terms of the count matrix $C_{ij}$,
\begin{equation}
M_{ij} = \frac{C_{ij}}{\sum_k C_{ik}} .
\end{equation}
Figure~\ref{fig:agreement_matrix} shows the agreement matrices computed from panel reviews in Cycles~0--7 for Stage~1 (left) and Stage~2 (right), using decile bins. For example, if Reviewer~A assigns a proposal to the top decile in Stage~1, there is a 19\% probability that Reviewer~B also assigns it to the top decile, but a non-negligible probability ($\sim$5\%) that Reviewer~B assigns it to the bottom decile. After Stage~2 discussion, agreement improves: the probability that both reviewers assign a proposal to the top decile increases to 34\%, though substantial dispersion remains. Overall, the agreement matrices demonstrate that panel discussion improves reviewer agreement, but does not eliminate significant variation in relative rankings.

\subsection{Simulating Rankings}

Using the agreement matrix defined in Section~\ref{subsec:ram}, we construct three simulation models that differ in their treatment of proposal-specific agreement levels:
\begin{itemize}
\item \textbf{Model 1}: Uses only the empirical agreement matrix to define the level of reviewer disagreement for a given proposal.
\item \textbf{Model 2}: Incorporates proposal-specific weighting to account for varying consensus levels in actual proposal reviews. Proposals are assigned random true quantiles.
\item \textbf{Model 3}: Same as Model 2, but with true proposal quantiles fixed to actual panel-derived rankings.
\end{itemize}

For all models, we use the proposals in Cycle~7 as the simulation basis, with proposals assigned to hypothetical DPR reviewers by randomly drawing from the pool of actual panel reviewers. A panel reviewer may be selected multiple times, with each selection treated as an independent reviewer assigned to a different proposal set. The assignment procedure follows \citet{Carpenter25}, except that reviewer--proposal similarity is set to unity only for reviewer--proposal pairs that occurred in the panel reviews.

In all simulations, each proposal is first assigned a latent underlying quality quantile. For a given simulated reviewer, the assigned quantile is drawn probabilistically from the corresponding row of the agreement matrix, conditioned on the proposal's true quantile. A small random jitter is added to break ties, after which reviewer-specific ranks are computed. Rank dispersion statistics (RMS and spread) are then calculated identically to the real DPR data (see Section~\ref{sec:ranks}).

\subsubsection{Model 1: Agreement Matrix Only}

\begin{figure}
\centering
\includegraphics[width=\columnwidth]{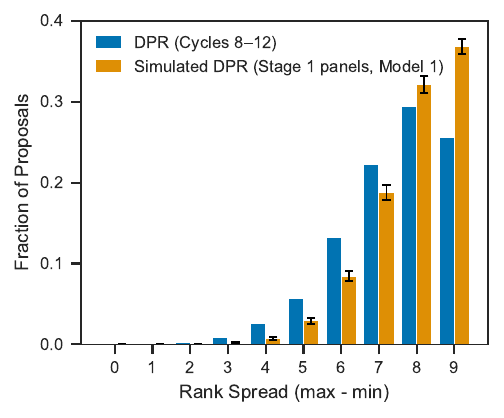}
\caption{Histogram of rank spread for observed Cycles~8--12 DPR proposals (blue) and simulated DPR rankings using Model~1 with Stage~1 panel agreement matrices (orange).}
\label{fig:spread_ranks_stage1_no_priors}
\end{figure}

Figure~\ref{fig:spread_ranks_stage1_no_priors} compares the observed rank spread in Cycles~8--12 DPR data with Model~1 simulations based solely on the agreement matrix. The simulation produces a larger fraction of proposals with maximal rank spread than observed, indicating that reviewer--reviewer agreement alone does not fully capture proposal-specific consensus.

\subsubsection{Model 2: Proposal-Specific Weighting with Random Quantiles}

Model~1 assumes that reviewer agreement depends only on a proposal's true quantile through the global agreement matrix. However, some proposals naturally exhibit stronger consensus among reviewers than others. To account for this proposal-specific variation, Model~2 applies a weighting factor based on the observed consensus for each individual proposal.

We define a raw consensus metric for proposal $p$ as
\begin{equation}
w_p^{\mathrm{raw}} = \frac{k_p}{n_p},
\end{equation}
where $k_p$ is the number of reviewers assigning proposal $p$ to its modal quantile bin and $n_p$ is the total number of reviews. This proportion can be noisy for proposals with a small number of reviews. To regularize the estimate, we apply Laplace smoothing and define the final weighting factor as
\begin{equation}
w_p = \frac{\alpha + k_p}{\alpha + \beta + n_p},
\end{equation}
with $\alpha = \beta = 1$, corresponding to a uniform $\mathrm{Beta}(1,1)$ prior on the underlying consensus probability. This choice introduces one pseudo-count for both agreement and disagreement, preventing weights of exactly 0 or 1 when few reviews are available, while having negligible impact for well-sampled proposals. The resulting weight smoothly interpolates between reliance on proposal-specific consensus when agreement is strong and reliance on the global agreement matrix when consensus is weak or poorly constrained. 

\begin{figure*}
\centering
\includegraphics[width=\textwidth]{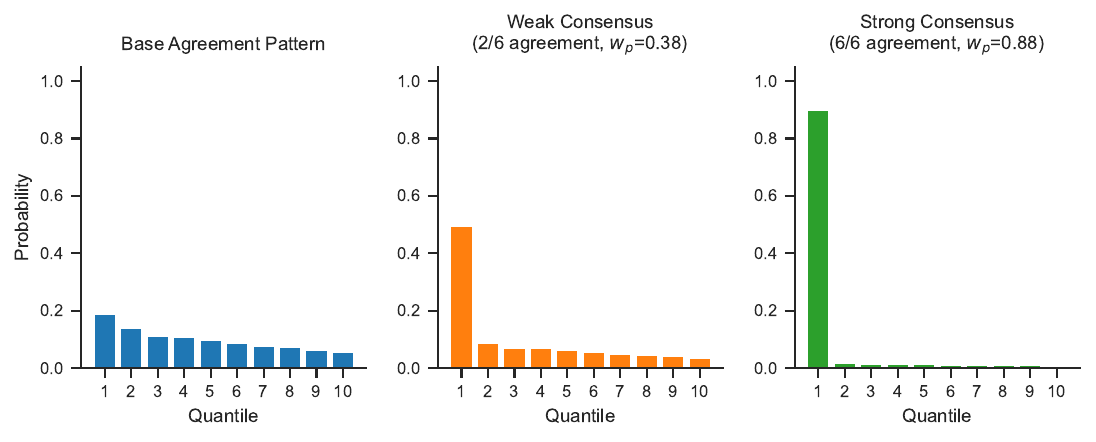}
\caption{Illustration of the proposal-specific weighting applied to the reviewer agreement matrix. Left: base agreement matrix. Middle: weak consensus (2 of 6 reviewers). Right: strong consensus (6 of 6 reviewers).}
\label{fig:weighting_agreement_function}
\end{figure*}

Figure~\ref{fig:weighting_agreement_function} illustrates the behavior of this weighting for representative weak- and strong-consensus cases. Including this weighting in Model~2 improves agreement between simulated and observed rank spreads, as shown in Figure~\ref{fig:spread_ranks_stage1}.

We also perform Model~2 simulations using the Stage~2 panel agreement matrix rather than Stage~1. These results are shown in Figure~\ref{fig:spread_ranks_stage2} and discussed in detail in the main text, where they are compared directly to the Cycles~8--12 DPR dispersion following the second ranking stage.

\subsubsection{Model 3: Proposal-Specific Weighting with Panel-Derived Quantiles}

\begin{figure}
\centering
\includegraphics[width=\columnwidth]{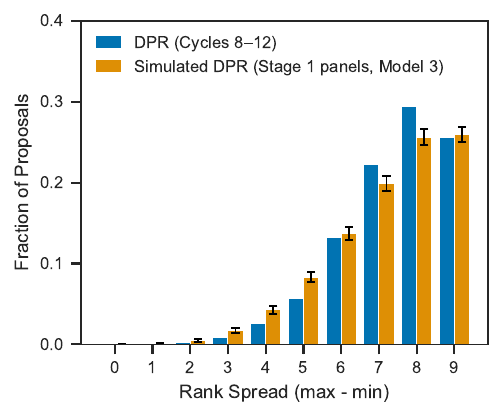}
\caption{Same as Figure~\ref{fig:spread_ranks_stage1_no_priors}, but using Model~3 with true proposal quantiles fixed to panel-derived rankings rather than assigned randomly.}
\label{fig:spread_ranks_stage1_priors_ranks}
\end{figure}

In the Model~2 simulations, proposals are initially assigned random true quantiles. As an alternative, Model~3 fixes the true quantile to the actual panel-derived rank. Results using this alternative assumption are shown in Figure~\ref{fig:spread_ranks_stage1_priors_ranks} and indicate that the largest rank spreads are slightly reduced relative to Model~2 but nonetheless remain similar to those observed in DPR.

\end{document}